\documentclass[aps,prl,twocolumn,groupedaddress,nofootinbib]{revtex4-1}

\usepackage[pdftex]{color,graphicx}
\usepackage{amssymb}
\usepackage{amsmath}
\usepackage{color}
\usepackage{array}
\usepackage{float}
\restylefloat{table}

\begin{document}


\title{One-dimensional photonic quasicrystals}


\author{Mher Ghulinyan}
\affiliation{Centre of Materials and Microsystems, Fondazione Bruno Kessler,  I-38123 Povo, Italy}


\date{Prepared on May 21st, 2013, corrected October 26th, 2014 \\
FFFrom: Light Localisation and Lasing, Eds. M. Ghulinyan, L. Pavesi, Cambridge Univ. Press (2015) Ch. 5, pp. 99-119}

\begin{abstract}
\end{abstract}


\maketitle
\tableofcontents
\section{I. Introduction}\label{intro}
\begin{figure}[b!]
\includegraphics[width=\columnwidth]{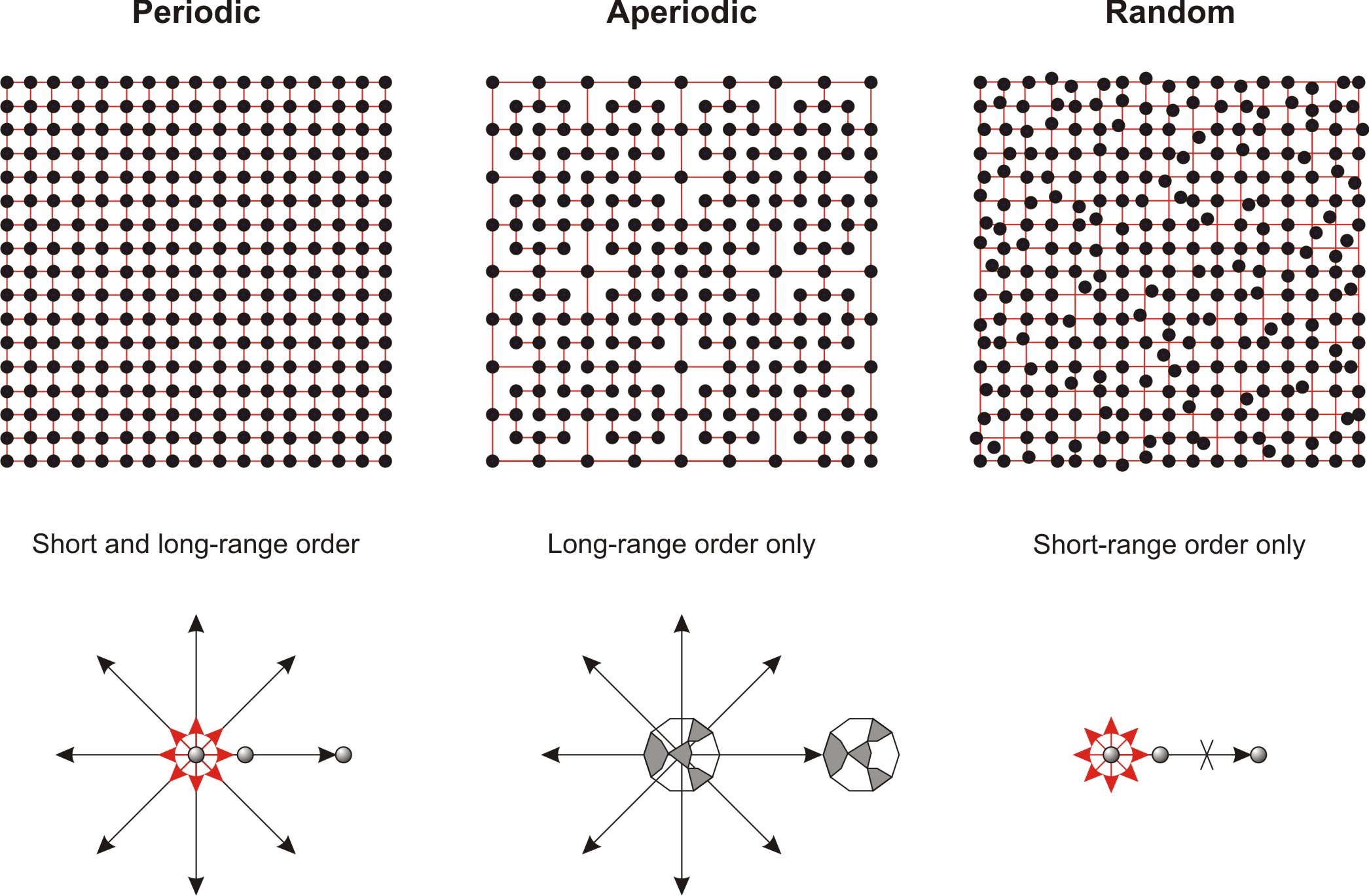}
\caption[From periodic to quasiperiodic and random]
{Schematic views of a two-dimensional (a) periodic, (b) a quasiperiodic and (c) a random lattice. Below each panel the corresponding situation for the atomic order is shown.}
\label{peraperand}
\end{figure}

\index{crystal! electronic}
Self organization is one of the most extraordinary tools in Nature which assembles its elementary building blocks, atoms and molecules, in the form of solid substances. Crystals, in this sense, represent the class of a vast number of solid materials in which atoms (molecules) are brought together in a perfect, spatially-periodic manner. A macroscopic crystal, is therefore made by repeating its smallest microscopic period -- the crystal \emph{unit cell} -- infinitely in space (Fig.~\ref{peraperand}(a)). Therefore, when translating the crystal by an integer number of unit cell size, the crystal coincides with itself. This property is in the basis of modern solid state physics and crystallography and is known as the \emph{translational symmetry}\index{symmetry! translational}. Crystals are thus ordered materials in which atoms show both short-\index{order! short-range} (within a couple of nearest neighbors) and long-range order\index{order! long-range}. Due to the periodic lattice\index{lattice! periodic} potential\index{periodic! potential}, the energy states within a crystal are extended, and the electrons can freely diffuse through it.

At the opposite extreme of crystalline order there are the disordered or amorphous solid materials, in which the short-range order may still preserve, while the long-range order misses completely (Fig.~\ref{peraperand}(c)). In these kind of solids, e.g. in heavily doped semiconductors or glasses, periodicity lacks and the atomic potential varies in a random manner. In certain cases, when the degree of randomness is large enough, electron wavefunctions\index{wavefunction} localize exponentially near individual atomic cites and the diffusion of charge carriers vanishes (Anderson localization) \cite{Anderson}\index{localization! Anderson}.

With the advances in $X$-ray diffraction tools in the beginning of the XX-th century, scientists learned to image the signatures of long-range order of crystals by mapping the symmetries of their reciprocal space lattices\index{lattice! reciprocal space}.
The crystal symmetries\index{crystal! symmetries} for all possible arrangements of perfect periodicity are classified in 230 space groups. On contrary, the diffraction\index{diffraction! pattern}\index{pattern! diffraction} form a disordered solid shows cloud-like concentric rings indicating to a diffraction from totally random atomic lattice and, consequently, a lack of any symmetry or, equivalently, any preferential direction in the solid.\\

\section{II. In between perfect periodicity and complete randomness}

In 1982 scientists found that certain ``crystalline" solids were showing point-like $X$-ray diffraction patterns with symmetries, e.g. 5-fold and 10-fold, that did not fit any of the known 230 space groups\footnote{According to the crystallographic restriction theorem the crystals are limited to 2-fold, 3-fold, 4-fold, and 6-fold rotational symmetries.} \cite{Shechtman}. These solids were named \emph{quasicrystals} and represent a large class of materials in between crystalline and disordered solids. A quasicrystal shows a perfect long-range order but does not possess a periodic lattice\index{lattice! periodic} (Fig.~\ref{peraperand}(b)). Equivalently, the quasicrystalline lattice can densely fill the space, however, it does not posses translational symmetry.

From a mathematical point of view, two- (2D) and three-dimensional (3D) quasicrystals are formed following aperiodic tiling\index{aperiodic! tiling}\index{tiling! aperiodic} rules. Discovered by mathematicians in early 60's\footnote{Girih tilings\index{tiling! Girih} were used in far 1200's for decoration of buildings in medieval Islamic architecture.}, the aperiodic tilings can fill densely the plane by a combination of scaling, rotating and repeating procedures of compound patterns. Similar rules allow to compound forms for filling the space \cite{2Dtylings}.

So far, on the way towards our understanding of the surrounding physical world and natural phenomena, scientists have adopted mathematical approaches which point to a reduction of an existing problem into a one-dimensional (1D) one.  This is because in 1D mathematical models can be described analytically and solved exactly. Higher-dimensional problems, in many cases, are then constructed based on a generalization of 1D models and their solutions, while in other cases, numerical methods and laborious computer simulations should run.

In the next section, we will address principal aspects of 1D quasiperiodicity with a particular focus on 1D Fibonacci chains. Further, the rest of the chapter will be dedicated to the electromagnetic counterpart of 1D Fibonacci structures as a relatively simplest case of the large class of photonic quasicrystals.

\section{III. 1D quasiperiodicity: Fibonacci chain}\label{fibochain}
\begin{figure}
\includegraphics[width=\columnwidth]{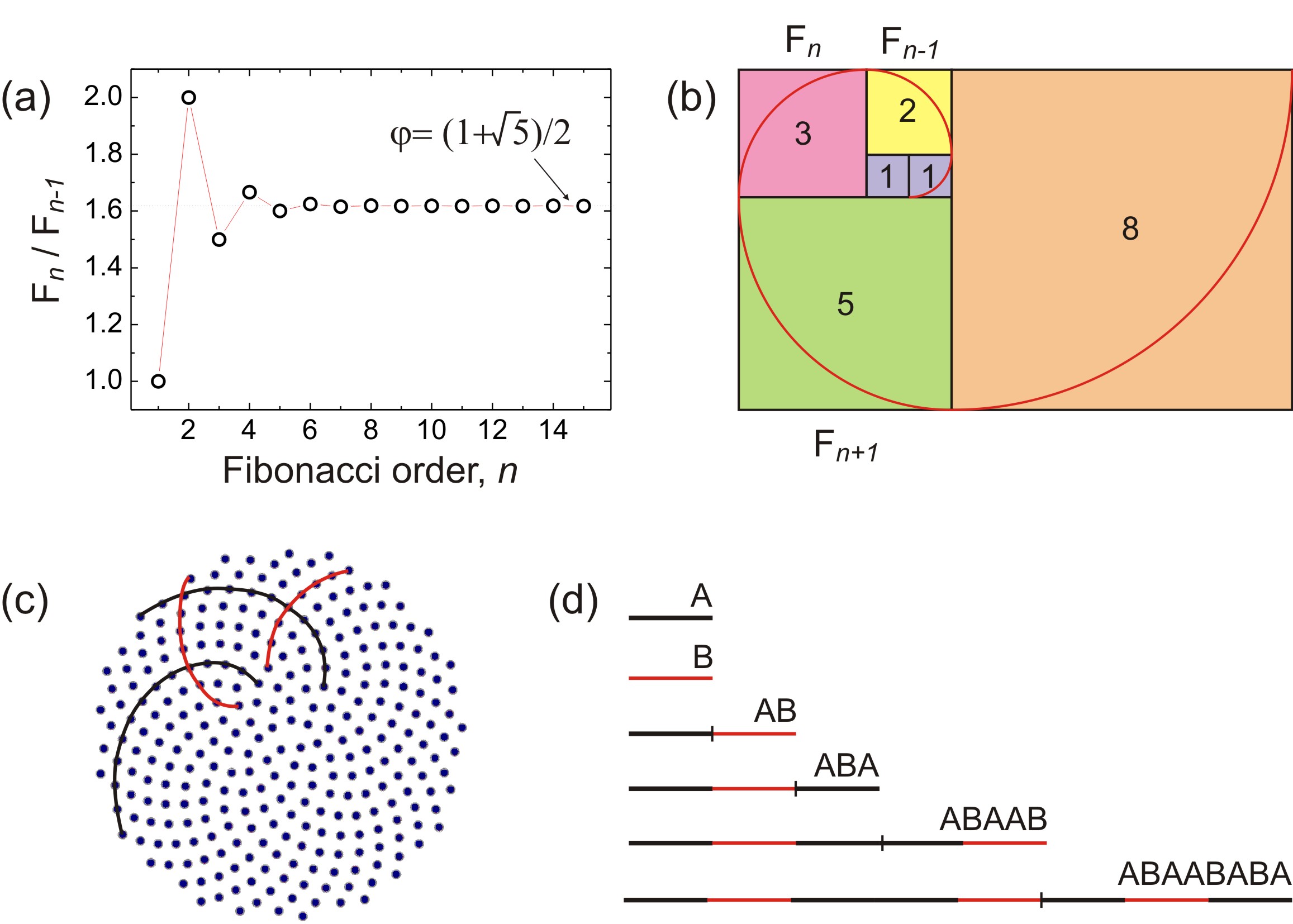}
\caption[Fibonacci recurrence rule]
{(a) The Fibonacci numbers and the golden ratio. (b) Tiling of squares with sides increasing as Fibonacci numbers and the golden spiral\index{golden! spiral}. (c) The spiral pattern of Vogel reproducing sunflower florets. (d) The two-component Fibonacci words.}
\label{fibovarie}
\end{figure}

\index{aperiodic! sequence}\index{Fibonacci! chain}\index{quasicrystal! Fibonacci}
Perhaps, the most well known and studied example of an aperiodic sequence of numbers is the Fibonacci sequence\index{Fibonacci! sequence} \cite{sigler}. The first two numbers, called the \emph{seed}, are two $1$'s and the following numbers
\begin{equation}
2,\,3,\,5,\,8,\,13,\,21,\,\ldots
\label{fibonums}
\end{equation}
are obtained by summing each time the last two of the list\footnote{In an alternative definition, the seed numbers are $0$ and $1$, but the recurrence rule is the same.}. Mathematically, the recurrence rule defines the $n$-th Fibonacci number\index{Fibonacci! numbers}\index{number! Fibonacci} as
\begin{equation}
F_n=F_{n-1}+F_{n-2}
\label{fiborec}
\end{equation}
with the seeds $F_0=1$ and $F_1=1$. The successive Fibonacci numbers, as shown in Fig.~\ref{fibovarie}(a), are intimately connected through the Golden ratio\index{golden! ratio}
\begin{equation}
\varphi=\lim_{n\rightarrow\infty}\frac{F_{n}}{F_{n-1}}=\frac{1+\sqrt{5}}{2}\approx1.61803\ldots
\label{gratio}
\end{equation}

The deterministic nature, that follows from recurrence rules used to create aperiodic sequences, and, in particular, that of Fibonacci type, allows to construct a variety of many different ``forms" -- strings or patterns. A nice visual example is the 2D tiling of squares\index{tiling! squares}, which are scaled such that their sides equal to Fibonacci numbers in length, as shown in Fig.~\ref{fibovarie}(b).

Scientists have observed that in many plants, such as sunflower and cactus, the florets and stickers are organized on composite spiral-like patterns following Fibonacci rules \cite{plants,phyllotactic}. A model to reproduce the sunflower floret pattern has been introduced by H. Vogel \cite{vogel}. It constitutes in generating a scatter plot in polar coordinates $\left(r=a\sqrt{n},\,\theta=2\pi n/\varphi^2\right)$, where $a$ is a scaling factor, $n$ is the index of the floret and $\varphi$ is the Golden ratio (Fig.~\ref{fibovarie}(c)). In this model, the angle $\theta_G=2\pi/\varphi^2=137.508^\circ$ is the so called Golden angle, \index{golden! angle} at which in a circle of circumference $a+b$ the ratio $a/b$ of arcs $a$ at $2\pi-\theta_G$ and $b$ at $\theta_G$ equals to $(a+b)/a$.

In 1D, by using recurrence rules, one may construct aperiodic sequences  starting from seed \emph{strings}. The most famous example is forming so-called ``aperiodic words"\index{aperiodic! word} by starting with seeds $A$ and $B$ and recurrently applying some rule for the next word-generations. In particular, in the case of a Fibonacci sequence\footnote{Another intensively studied aperiodic sequence is the Thue-Morse one which uses an \{$A\rightarrow AB$, $B\rightarrow BA$\} inflation \cite{Thue}.}, an inflation rule is used
\begin{eqnarray}
  A &\rightarrow& AB \\
  B &\rightarrow& B.
\end{eqnarray}
The first six Fibonacci strings, therefore, are $A$, $B$, $AB$, $ABA$, $ABAAB$ and $ABAABABA$ (Fig.~\ref{fibovarie}(d)). As it is seen, this inflation rule is equivalent to adding the $(n-2)$th sequence to the end of the $(n-1)$th in order to generate the $n$th series, $F_n=F_{n-1}F_{n-2}$. One particularity of the Fibonacci string is that, while $AA$ is a frequent instance along an $F_n$ for large $n$'s, other instances such as $BB$ or $AAA$ may never appear. As it will be discussed in the next section, this particularity appears to have significance in the context of a 1D photonic Fibonacci-type quasicrystal.

\begin{table}[t!]
\label{sample-table}
\addtolength\tabcolsep{2pt}
\begin{tabular}{@{}c@{\hspace{25pt}}ccc@{}}
\hline \hline
\raggedleft
String & Length & counts($A$) & counts($B$)\\
\hline
\multicolumn{1}{|l|}{$A$} & 1
& 1 & 0\\[3pt]
\multicolumn{1}{|l|}{$B$} & 1 & 0 & 1\\[3pt]
\multicolumn{1}{|l|}{$AB$} & 2 & 1 & 1\\[3pt]
\multicolumn{1}{|l|}{$ABA$} & 3 & 2 & 1 \\[3pt]
\multicolumn{1}{|l|}{$ABAAB$} & 5 & 3 & 2 \\[3pt]
\multicolumn{1}{|l|}{$ABAABABA$} & 8 & 5 & 3 \\[3pt]
\multicolumn{1}{|l|}{$ABAABABAABAAB$} & 13 & 8 & 5 \\[3pt]
\hline \hline
\end{tabular}
\caption[Fibonacci strings] {\small{The Fibonacci strings and relations between its length and counts of components. Length/counts($A$)$\rightarrow\varphi$ and counts($A$)/counts($B$)$\rightarrow\varphi$ for long Strings.}}
\end{table}

\section{IV. Photons in a 1D optical potential}\label{1dphot}\index{crystal! photonic}
The energy spectrum of a quantum particle in a semiconductor crystal is described by extended Bloch modes and consists of allowed and forbidden bands. Electronic crystals have their analogue in the world of electromagnetic waves (photons) called \emph{photonic crystals} \cite{PhCryst}. Photonic crystals are complex materials, in which a periodic variation of the material dielectric constant, $\epsilon(\vec{r})$, on a length scale comparable with the wavelength leads to the formation of energy bands where the propagation of photons is allowed or forbidden\index{bandgap! photonic}.

\begin{figure}
\includegraphics[width=\columnwidth]{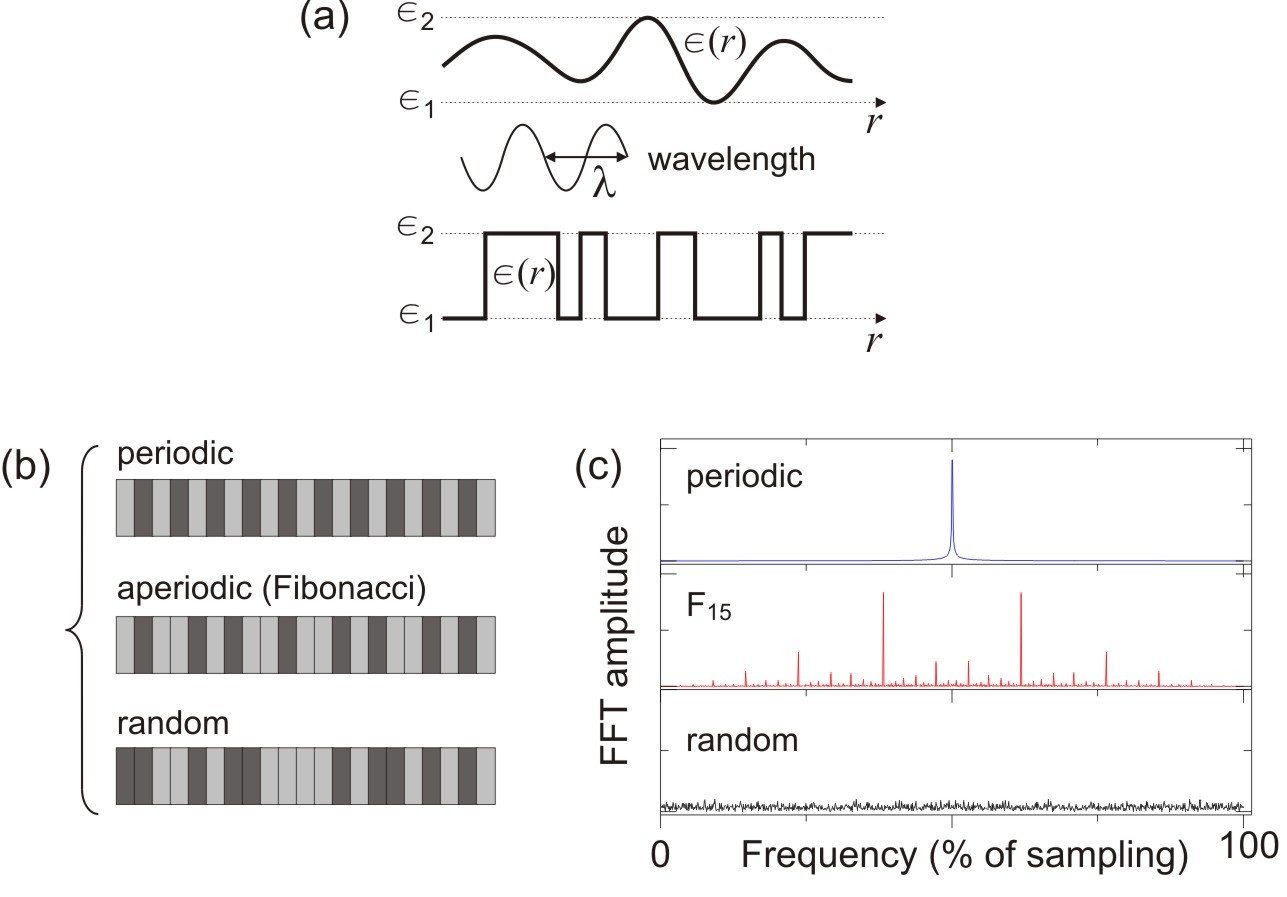}
\caption[A complex dielectric medium]
{(a) The variation of the dielectric constant in space on a length scale comparable to the light wavelength lays in the basis of photonic crystals. $\epsilon(\vec{r})$ may vary either in a smooth or a step like manner. (b) With a step-like change of $\epsilon(\vec{r})$, by using to different materials $A$ and $B$, results in the formation of a periodic (top), a quasiperiodic (middle) or a completely random (bottom) optical potential for photons. (c) The  Fourier transformation spectra of the corresponding bi-component layer sequences show a single sampling frequency (periodic), a self-similar frequency distribution pattern (quasiperiodic) and an uncorrelated flat spectrum with no any specific frequency (random).}
\label{1Dphotonic}
\end{figure}

Likewise solid materials, photonic crystals can be periodic, quasiperiodic and random, reflecting the variation of $\epsilon(\vec{r})$ in space. This variation, in turn, can be either continuous (Fig.~\ref{1Dphotonic}a, top) or abrupt with the space coordinate (Fig.~\ref{1Dphotonic}(a), bottom). In the first case a smooth variation of $\epsilon(\vec{r})$ between $\epsilon_1$ and $\epsilon_2$ may reflect a continuous change in material composition (an Al$_x$Ga$_{1-x}$As alloy, for example). In the second case an abrupt change in $\epsilon(\vec{r})$ is achieved when two different materials with dielectric constants $\epsilon_1$ and $\epsilon_2$ are stack together.

Throughout this chapter, we will focus on the optical properties of 1D photonic quasicrystals of Fibonacci type which are formed by stacking two different materials, say $A$ and $B$, following the inflation rule discussed above. Prior to this, we will introduce the basic properties of 1D photonic multilayered structures on the simple examples of a Distributed Bragg Reflector\index{Bragg! reflector} (DBR) and a Fabry-P$\acute{e}$rot microcavity.\\

\subsection{Distributed Bragg Reflector}
\begin{figure}[t!]
\includegraphics[width=\columnwidth]{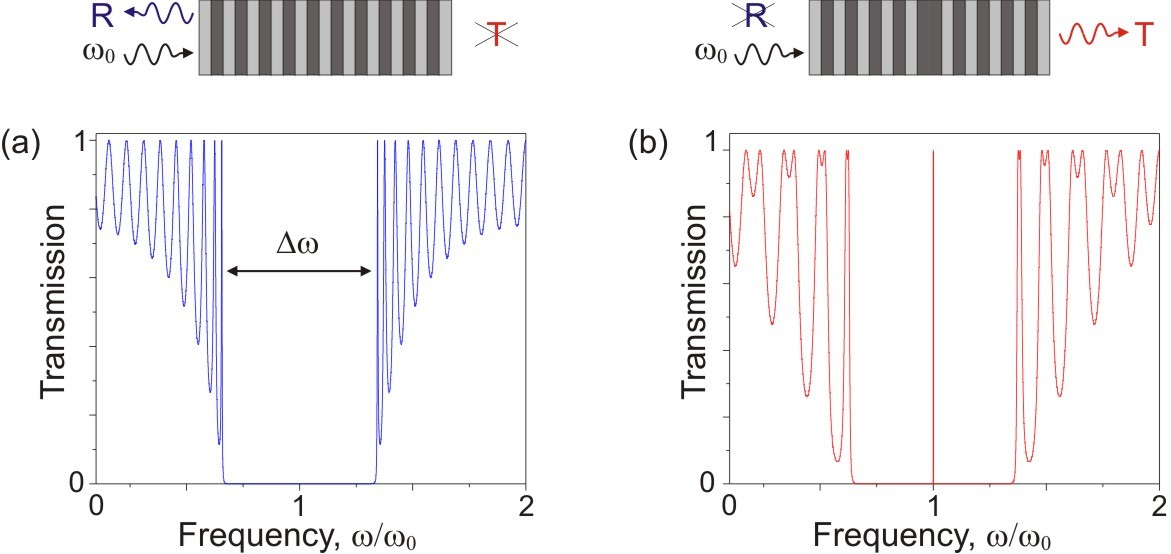}
\caption[Elementary 1D photonic crystals: Transmission spectra]
{The multilayer structure and the transmission spectra of (a) a Distributed Bragg Reflector and (b) a Fabry-P$\acute{e}$rot microcavity.}
\label{DBR}
\end{figure}

Otherwise known as a Dielectric Mirror, the functionality of a DBR is based on the interference\index{interference! of waves} between electromagnetic waves\index{wave! interference} (EMW) which are multiply reflected and refracted at each boundary of the multilayer structure. An EMW of frequency $\omega_0$ is efficiently reflected\footnote{For simplicity, we will consider the case of a normal incidence in the following.} from a DBR if the multilayer is constructed by alternating quarter-wave layers of two different materials $A$ and $B$ with refractive indices $n_A=\sqrt\epsilon_A$ and $n_B=\sqrt\epsilon_B$, respectively. The quarter-wave condition implies that $n_A d_A=n_B d_B=\lambda_0/4$, where $d_A$ and $d_B$ are the layer thicknesses, $\lambda_0=2\pi c/\omega_0$ is the wavelength  and $c$ is the EMW velocity in free space. The portion of the EMW intensity reflected from the multilayer embedded in air from both sides is given by
\begin{equation}
R=\left[\frac{n_A^{2N}-n_B^{2N}}{n_A^{2N}+n_B^{2N}}\right]^2,
\label{dbr_R}
\end{equation}
where $N$ is the number of periods ($AB$ pairs) of the DBR. In absence of losses (scattering, absorption etc.) the transmission of the mirror is $T=1-R$. As shown in the example of Fig.~\ref{DBR}(a), the transmission vanishes within a bandwidth\index{bandwidth! spectral} of $\Delta\omega$ around the frequency $\omega_0$, which means that in this frequency window -- \emph{the photonic band gap} -- EMWs cannot propagate through the structure. The relative width of the photonic band gap is  given approximately by
\begin{equation}
\frac{\Delta\omega}{\omega_0}=\frac{4}{\pi}\arcsin\left(\frac{|n_A-n_B|}{n_A+n_B}\right).
\label{dbr_gapwidth}
\end{equation}

From Eq.~(\ref{dbr_R}) we see that the reflectivity of the dielectric mirror grows with either increasing the number of periods $N$ or increasing the refractive index contrast $\frac{|n_A-n_B|}{n_A}$. This last is also crucial for making a wider photonic band gap, as follows from Eq.~(\ref{dbr_gapwidth}), while increasing $N$ affects only the steepness of the band edges\index{bandgap! edge}.

A common mathematical approach to calculate the spectral properties of 1D multilayered photonic structures is based on the transfer-matrix method\index{method! transfer-matrix}\index{transfer! -matrix} \cite{Pedrotti-TMM}. This method consists in relating the electromagnetic fields on two boundaries (left and right, hereafter) of the dielectric layer by means of the transfer matrix, $T$. In the case of the electric filed $E$ this relation is
\begin{equation}
E_{l}=M\times E_{r}=\left( \begin{array}{cc} \cos{\delta} & i \frac{c}{n}\sin{\delta} \\ i\frac{n}{c}\sin{\delta} & \cos{\delta} \end{array} \right)E_{r},
\label{trmatrix}
\end{equation}
where $n$, $d$ and $\delta=\frac{2\pi n d}{\lambda}$ are the refractive index, the thickness of the layer and the phase change across it, respectively, $E_{l,r}$ are the electric fields at the left and right boundaries of the layer. Equation~(\ref{trmatrix}) in the same form holds also for the magnetic field.

From continuity requirements it follows that the transfer matrix $M$ for a system of $N$ layers is obtained by simple matrix multiplication of $M$'s of single layers, $M=\prod M_j$ with $M_j$ being the transfer matrix of the $j$th layer. The complex reflectance, $r(\omega)$ and transmittance, $t(\omega)$ coefficients of the multilayer can be thus calculated from the total transfer matrix $M$ by assuming a unity (normalized to the input) intensity impinging on the structure from one side (left) and no input from the other (right):
\begin{equation}
r(\omega)=\frac{\gamma_l m_{11}+\gamma_l\gamma_r m_{12}-m_{21}-\gamma_r m_{22}}{\gamma_l m_{11}+\gamma_l\gamma_r m_{12}+m_{21}+\gamma_r m_{22}},
\label{cmplx_r}
\end{equation}
\begin{equation}
t(\omega)=\frac{2 \gamma_l}{\gamma_l m_{11}+\gamma_l\gamma_r m_{12}+m_{21}+\gamma_r m_{22}},
\label{cmplx_t}
\end{equation}
where $\gamma_l$ and $\gamma_m$ are the inverse of light velocities in the input and output media, respectively.

\begin{figure}[b!]
\includegraphics[width=\columnwidth]{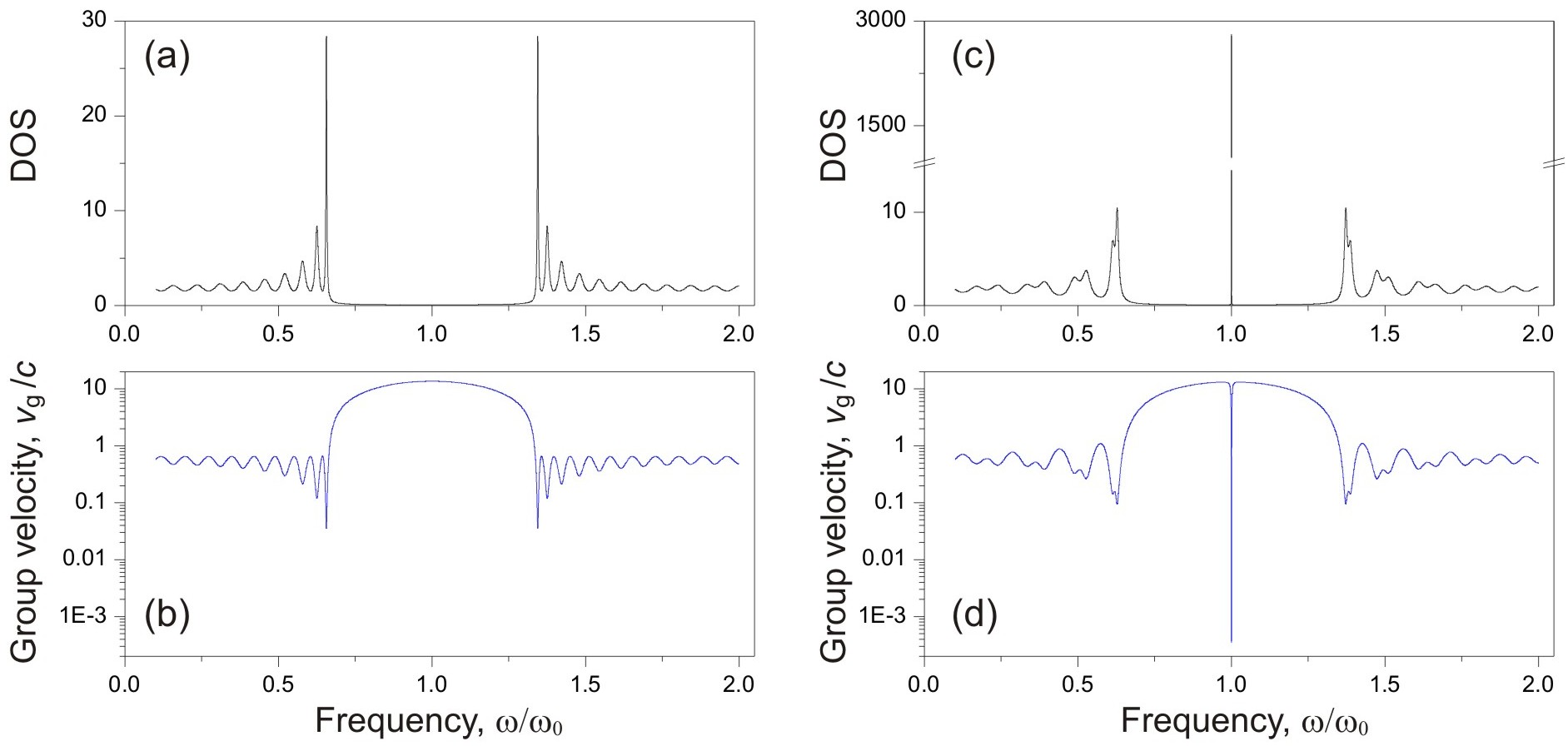}
\caption[Elementary 1D photonic crystals: DOS and group velocity]
{The density of states (DOS) and the group velocity for the cases of (a) a dielectric mirror DBR and (b) a Fabry-P$\acute{e}$rot microcavity. In the last case a strong enhancement of the DOS and significantly reduced group velocity can be observed at the resonant frequency $\omega/\omega_0=1$.}
\label{dom}
\end{figure}

The complex quantities contain important information about the photonic properties of a multilayer. For example, the power reflection and transmission coefficients are calculated as $R=|r(\omega)|^2$ and $T=|t(\omega)|^2$, respectively. Moreover, the amplitude and the phase of the transmitted light can be retrieved from $t(\omega)$. The phase $\phi$, in particular, which is the argument of the complex transmittance, $\arg[t(\omega)]$, allows one to calculate both the density of states (DOS)\index{density! of states} of the photonic structure, $\rho(\omega)$, and the group velocity of propagating light\index{group velocity}, $v_g(\omega)$, as
\begin{equation}
\rho(\omega)=v_g(\omega)^{-1}=L^{-1}_{tot} \frac{d\phi}{d\omega},
\label{DOS}
\end{equation}
with $L_{tot}$ being the total physical thickness of the multilayer. Figure~\ref{dom}(a) plots the DOS of the same dielectric mirror of Fig.~\ref{DBR}(a). It shows, in particular, that the DOS, defined typically as the number of states (photonic modes in this case) per unit energetic (frequency) interval, is largely enhanced at certain frequencies adjacent to photonic band gap, known as \emph{band edge states}.

Following the definition of Eq.~(\ref{DOS}), the group velocity of a wavepacket\index{wavepacket} centered at the band edge frequencies and propagating through the multilayer is strongly reduced for those spectral components which lay within the narrow high-DOS lines (see Fig.~\ref{dom}(b)). As a consequence, the temporal evolution of a transmitted through DBR light pulse shows important wavepacket broadening as a result of a pronounced delay of these spectral components (mode localization).\\

\subsection{Fabry-P$\acute{e}$rot microcavity}\index{microcavity! Fabry-P$\acute{e}$rot}
The Fabry-P$\acute{e}$rot (FP) microcavity is a dielectric mirror with a defect. The simplest way to introduce a defect in a dielectric multilayer is to eliminate one of the $\lambda_0/4$ layers of either type $A$ or $B$ materials. By doing this, two quarter-wave layers of the same material appear next to each other, forming thus a $\lambda_0/2$ layer which leads to a perfect constructive interference\index{interference! constructive} of transmitted at $\omega_o$ EMW. As a result, a sharp (unity) transmission channel\index{channel! transmission} is formed at the center of the photonic band gap (Fig.~\ref{DBR}(b)). The narrowness of the transmission line depends on the reflectivity (i.e., the strength) of the dielectric mirrors sandwiching the half-wavelength defect layer. As for the reflectance of the FP microcavity, it vanishes exactly at the resonant frequency, while for the rest of frequencies shows all the spectral features of a DBR.

It is important to note the dramatic increase of the DOS (Fig.~\ref{dom}(c)) and, consequently, the suppression of group velocity at the microcavity resonance frequency (Fig.~\ref{dom}(d)). These are signatures of a strongly localized defect state for which the mode wavefunction decays exponentially outside the defect. The ability to confine EMWs in a tiny volume such as that of a defect layer, makes microresonator\index{resonator! micro-} devices excellent tools for  studying light-matter interactions at a nanometric scale. Lasing and Purcell enhancement \index{Purcell! enhancement} of the radiative rates of cavity-embedded emitters are typical phenomena which largely exploit the confining properties of microresonators.

Finally, after the introduction to the physics of a dielectric mirror and a microcavity, we now have all the basic ingredients necessary to start with the optical properties of a 1D photonic quasicrystal. In the last part of this chapter we will see how a 1D Fibonacci quasicrystal is formed and in which way the quasiperiodic order\index{order! quasiperiodic} modifies the optical properties of an otherwise ordered photonic structure.

\section{V. Photons in a 1D quasiperiodic potential}\label{1dquasi}\index{quasicrystal! photonic}

As briefly cited in the beginning of the previous section, an optical potential for electromagnetic waves can be realized by properly stacking thin quarter-wave layers made of two different dielectric materials $A$ and $B$. In the two extremes, a strictly periodic\index{periodic! structure} alternation of these layers will form a perfectly ordered 1D photonic crystal (DBR), while a complete randomization of the stacking order will give rise to a disordered photonic structure (Fig.~\ref{1Dphotonic}(b)).

\begin{figure}
\includegraphics[width=\columnwidth]{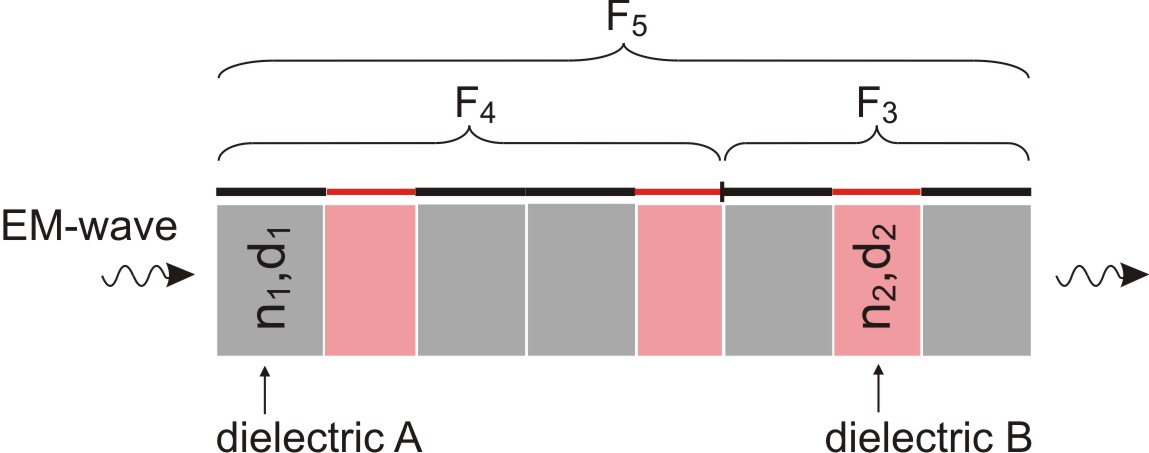}
\caption[The 1D photonic Fibonacci quasicrystal]
{A photonic quasiperiodic potential for electromagnetic waves is formed by alternating according the Fibonacci inflation rule two different dielectric layers of appropriately chosen optical thicknesses $n_i d_i$.}
\label{modelFQ}
\end{figure}

In terms of the positional order\index{order! positional} of $A$ and $B$ layers, the a 1D quasicrystal stands in between the perfect DBR and a random photonic structure. As already introduced in the Section~\ref{1dphot}, following the aperiodic string\index{aperiodic! string} generation rule, we now can construct the photonic quasicrystal structure of Fibonacci type (FQ, hereafter) in which the constituent $\lambda/4$ layers of  $A$ and $B$-type materials are stack together obeying $F_n=F_{n-1}F_{n-2}$ for the $n$th Fibonacci-generation \cite{Gellermann,Hattori}. Thus, the number of $\lambda/4$ layers of an $n$th-generation FQ will correspond to the $n$th-order Fibonacci number (see Fig.~\ref{modelFQ}).

The differences between the periodic, aperiodic and completely random arrangements of two-component 1D systems of the same length (all $A$'s and $B$'s together) can be recognized by looking at the Fourier transform spectra of these sequences. Figure~\ref{1Dphotonic}(c) reports such spectra for a DBR, a 15th order FQ and a positional-random\index{random! positional} multilayer stack, all containing 987 layers. We see immediately that in the case of the DBR (top panel) a single occurrence frequency is present, indicating the perfect alternation of $A$ and $B$ layers. Instead, the flat spectrum of the random structure (bottom panel) is a strong evidence that no any correlation is present in the sequence of alternation of layer types. More interesting and rich is the Fourier transformation spectrum of the FQ (middle panel). We notice that this $A$/$B$-appearance spectrum represents a self-similar structure full of intense and weaker peaks -- signatures of long-range order and absence of perfect short-range periodicity.

\subsection{Electronic energy spectrum of 1D FQ}\index{quasicrystal! electronic}
Various authors have studied the energy spectrum and the properties of 1D quasicrystals. Early studies were mainly focused on electronic quasicrystals and much work has been done there. In particular, it has been reported that the typical structure of the energy spectrum, $T(\omega)$, of a Fibonacci system has a self-similar nature \cite{Kohmoto,Kohmoto2}. More technically, it is a Cantor set with zero Lebesgue measure \cite{Kohmoto}. In other words, an arbitrary chosen frequency lays within the gap with a unity probability, and the numerous gaps are densely filling the spectrum. In these spectral gaps, known as \emph{pseudo band gaps}\index{bandgap! pseudo-}, the DOS tends to zero \cite{Nori,Capaz}.

It was further shown that as a consequence of the multi-fractal nature of the energy spectrum, the wavefunctions\index{wavefunction} of a Fibonacci quasicrystal\index{Fibonacci! quasicrystal} show certain exotic localization properties \cite{Kohmoto2}. Namely, the wavefunctions are neither localized nor extended -- is is said they are \emph{critically localized}\index{localization! critical} -- and thus decay at a rate slower than an exponential one \cite{Soukoulis,Kohmoto,Fujiwara}. These statements have been multiply confirmed through numerical calculations and experiments \cite{Sokolof,Wang,Macia,Piechon,Huang,Steinbach}. Starting from mid 90's different studies with their optical (photonic) counterparts confirmed major analogies to the electronic case \cite{Bendickson,Gellermann}.

\subsection{Energy spectrum of 1D photonic FQ}
\emph{From electronic to photonic} -- Interference is a wave phenomenon \cite{ping}. The wave nature of both electrons and photons is the basis of important analogies between electronic charge transport \index{transport! of electrons} phenomena in solids and light propagation in complex dielectric systems \cite{skipetrov}. The spectrum of these analogies includes but is not limited to phenomena such as optical counterpart of weak localization \cite{Kuga,Albada1,18}, Anderson localization \cite{Dalichaouch,Genack,33}, short and long range correlations \cite{Garcia,Albada2}, and universal conductance fluctuations \cite{Scheffold}, photonic Bloch oscillations \cite{Sapienza} and Zener tunneling of light waves \cite{Ghulinyan1}.

Since photons are uncharged particles and interact extremely weakly with each other at relatively low light intensities, a light wave packet, that propagates through a dielectric system, remains coherent for much longer times than charged particles. This means that dynamic interference effects could be isolated and studied more easily with light than with electrons. In view of this, 1D photonic Fibonacci quasicrystals were not only realized and spectrally characterized in static experiments, but more importantly, photon wavepacket propagation through optical states around the pseudo band gap of the structure was studied thoroughly \cite{Luca,MherFibo}.\\

\emph{Transmission spectrum} -- In general, the energy spectrum, $E(\omega)$, of a photonic structure is often identified to its transmission spectrum $T(\omega)$. We will use this last definition throughout the rest of the chapter. As mentioned in the Section~\ref{1dphot}, the transmission spectrum of a multilayered FQ can be easily calculated according the transfer-matrix formalism (Eqs.~(\ref{trmatrix}) and (\ref{cmplx_t})).

\begin{figure}
\includegraphics[width=\columnwidth]{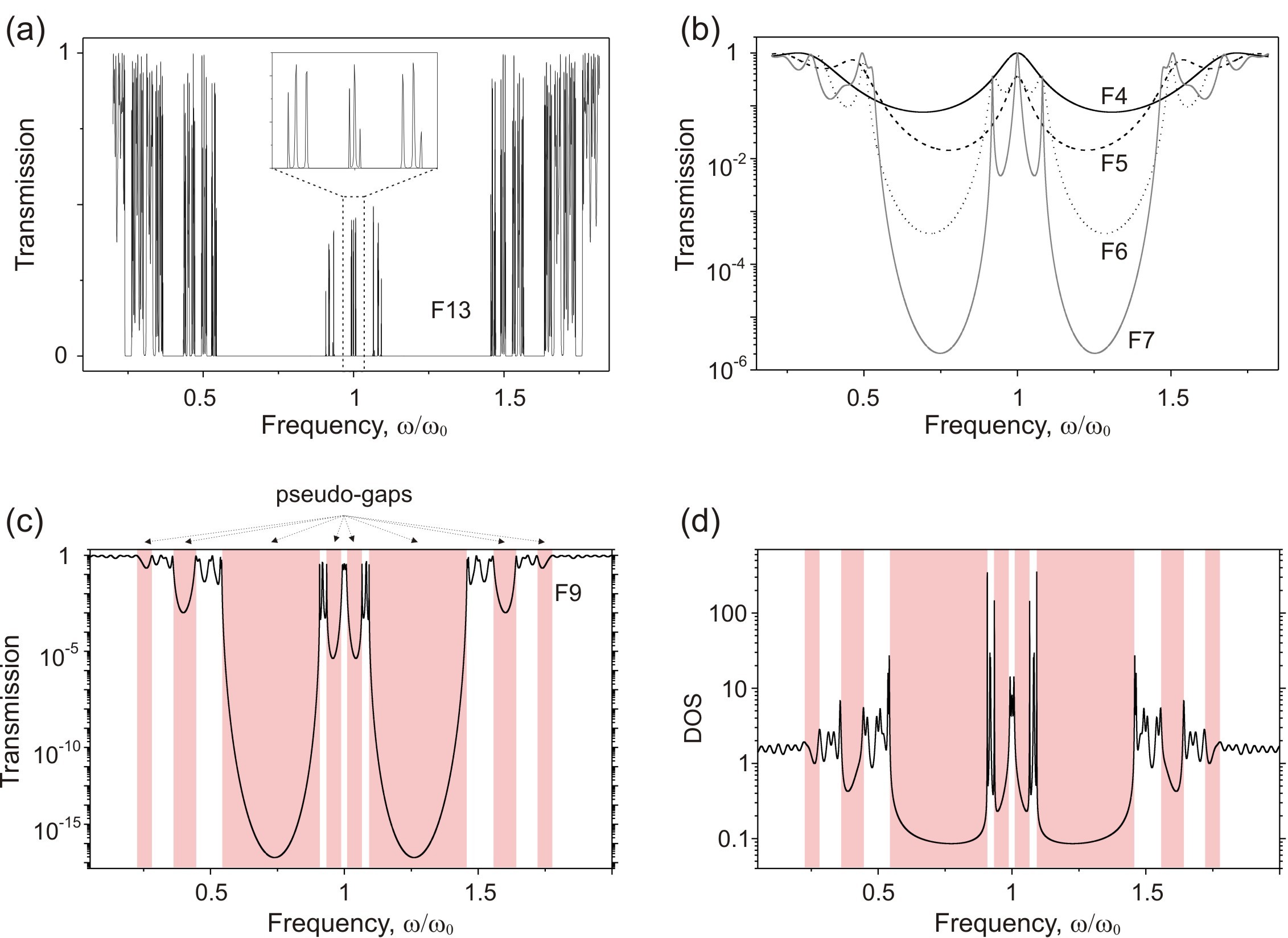}
\caption[Photonic modes in 1D FQs]
{(a) The transmission spectrum of a 13th order FQ with 377 layers. The inset shows the zoom around the central frequency revealing the fine modal features. (b) The evolution of the transmission spectra with increasing Fibonacci order: the pronounced deepening of low-$T$ regions is accompanied by the appearance of a richer resonant structure. (c) The shadowed regions in the spectrum of an F9 structure indicate the so-called pseudo band gaps. (d) The same is illustrated on the DOS spectrum. }
\label{F9}
\end{figure}

Figure~\ref{F9}(a) shows the transmission spectrum of a 377-layer 13th-order Fibonacci sequence with chosen refractive indices of $n_A=3$ and $n_B=1$. This spectrum is different form that of a perfect DBR but it represents a mixture of general spectral features of a dielectric mirror and a FP-microcavity in terms of clearly visible gap regions and narrow transmission lines. The calculated $T(\omega)$-spectra for four successive Fibonacci series are plotted in Fig.~\ref{F9}(b). The logarithmic scale of transmission shows how the spectra are evolved from $F_4$ to $F_7$ reflecting the deepening of prohibited spectral regions and the appearance of new narrow spectral features.\\

\subsection{Experiments with 1D photonic FQ's}
One of the first realizations of 1D FQ's and the spectral characterization of samples has been reported by Gellermann \emph{et al.} in 1994 \cite{Gellermann}. An electron-gun evaporation technique was used to grow up to 55 layer (F9) samples composed by silica and titanium dioxide layers alternating following the Fibonacci rule. The quarter-wave ($\lambda_0/4$) layers were centered a wavelength of $\lambda_0=700$~nm and static transmission spectra of the various Fibonacci samples were recorded around $\lambda_0$. The experimental results were successfully compared to the theory based on the transfer-matrix approach. As a typical drawback from the control over growth parameters, in particular, a variation of layer refractive index with increasing the structure size, certain deviation from the nominal optical path ($n_i\times d_i$) was observed too. By fitting the experimental experimental spectrum a 5\% of refractive index drift was estimated by the authors for F9 samples.

In works by Dal Negro \emph{et al.} \cite{Luca} and Ghulinyan \emph{et. al.} \cite{MherFibo} dielectric Fibonacci quasicrystals  were realized by electrochemical etching of silicon. Porous silicon has interesting optical properties \cite{Cullis} and offers an excellent platform for testing ideas experimentally in 1D multilayered photonic structures \cite{Luca,MherAPL,MherJAP,Sapienza,Ghulinyan1,MherFibo}. When grown in highly doped p-type silicon, porous silicon behaves as an optically homogeneous dielectric material with an effective refractive index $n$ determined by its porosity. Since it is fabricated by a self-limiting
electrochemical process, one can control the refractive index
of each layer by varying the electrochemical current during
the fabrication. Typically, this last follows a step variation between two constant values, and therefore results in a step-like variation of the effective refractive index profile through the whole structure (the bottom panel case of Fig.~\ref{1Dphotonic}(a)). The thickness $d$ of an individual porous layer is then determined by the etching duration under a constant current density. Hence, the control of etching current and the time guarantee a careful control over the optical path throughout the multilayered sample.

This growth technique was used successfully to realize porous silicon-based 1D FQ's on silicon substrates and to characterize them both in static transmission and ultrashort-pulse interferometric experiments at near-infrared (NIR) wavelengths \cite{Luca}. In particular, mode beating, sizable field enhancement, strong pulse stretching, and a strongly reduced group velocity in the band edge region around a Fibonacci pseudo band gap were all observed and interpreted in terms of Fibonacci band-edge resonances.

\begin{figure}
\includegraphics[width=\columnwidth]{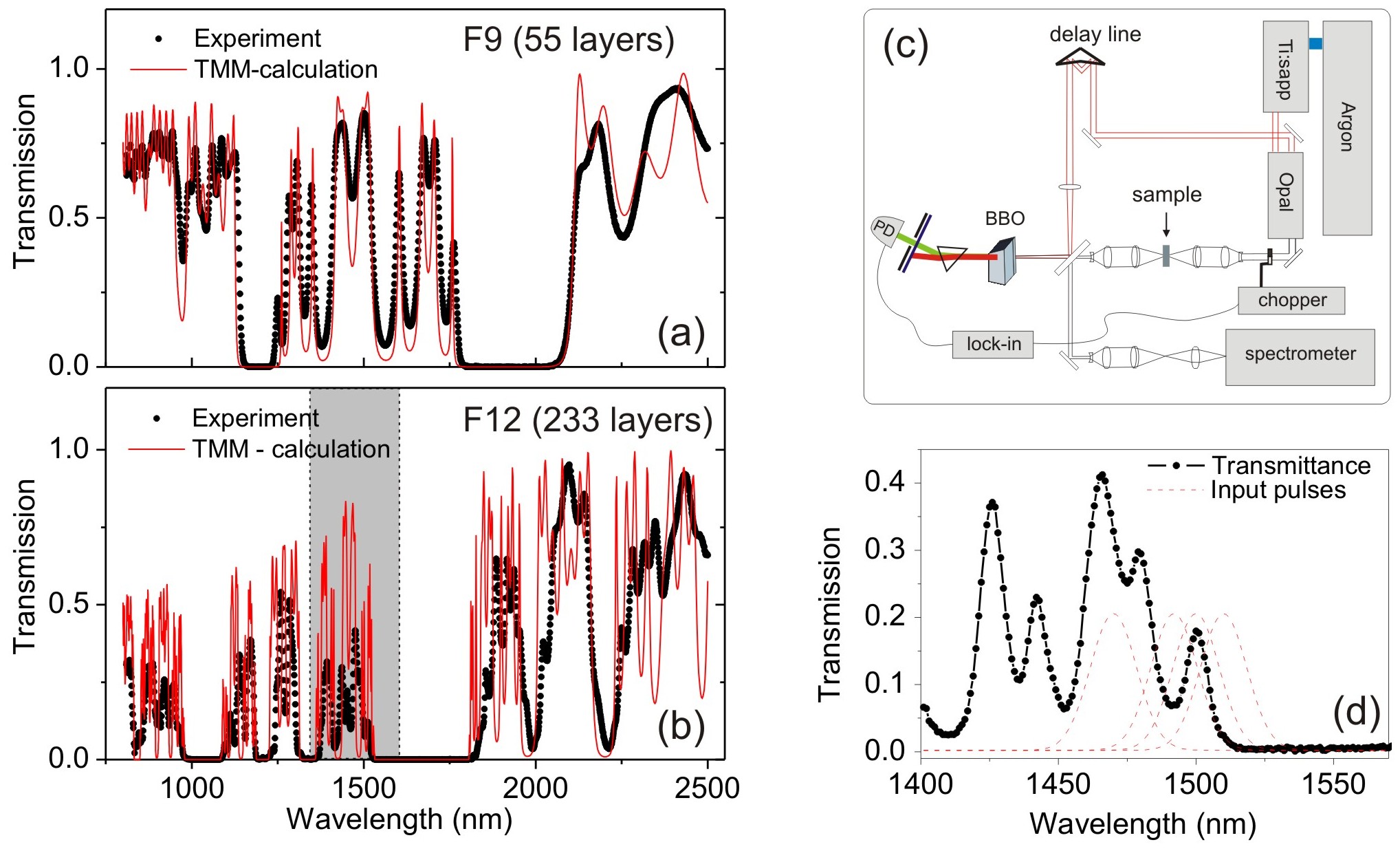}
\caption[Experiments with 1D FQs]
{(a) The transmission spectrum of the porous silicon-based F9 sample is compared to the transfer-matrix calculations with 1\% of optical path gradient. (b) The same for the F12 sample. The gradient is of 4\% in this case. (c) The ultrafast pulse transmission setup based on an optical gating technique. (d) Pulse propagation experiments were performed over the spectral range, which corresponds to the shaded area of panel (b). Dotted lines show examples of Gaussian profiles of the ultrashort pulses centered at different wavelengths \cite{MherFibo}. }
\label{prb1}
\end{figure}

A step forward in this direction was preformed by further improving the electrochemical growth technique \cite{MherFibo}, which allowed in particular to realize free-standing multilayered FQ's (without Si substrate). By using free-standing structures, in general, natural drifts of the nominal optical path can be controlled at a higher precision permitting to realize thick structures with high optical quality \cite{MherAPL,MherJAP}. Moreover, with such samples not only simpler static transmission experiments become accessible but as well one avoids possible large time-offsets due to the silicon substrate during pulse propagation studies. Along with the observation of all the various effects reported in the previous study \cite{Luca}, experiments with the new samples provided a further insight into the physics of photonic 1D FQ's.

In Figs.~\ref{prb1}(a),(b) the measured transmission spectra of F9 and F12 (233 layers) are shown. The transfer-matrix calculations, considering the presence of possible drifts in growth parameters, revealed a 1\% and 4\% of optical path gradient (linear to first approximation) for F9 and F12 samples, respectively (solid lines Figs.~\ref{prb1}(a),(b)). The central wavelength of the F9 sample was chosen to be approximately 1500~nm's. Instead, the sample F12 was intentionally realized such that the edge states of one of the pseudo band gaps are spectrally positioned around 1500~nm (see the shaded area in Fig.~\ref{prb1}(b)), where ultrashort pulse propagation studies could be performed (available range for sweeping the pulse wavelength).\\

\emph{Ultrashort pulse propagation through 1D FQ's} -- Ultra-fast time-resolved transmission measurements were performed using an
optical gating technique based on signal upconversion (Fig.~\ref{prb1}(c)). While being simpler compared to an interferometric system \cite{Luca}, it is sensitive to the transmitted intensity only and not the phase. In experiments wavelength-tunable pulses between 1400 to 1570 nm and of 220 fs duration (bandwidth\index{bandwidth! pulse} of 14~nm) were sent through the F12 sample probing thus the band edge states (Fig.~\ref{prb1}(d)). The transmitted signal was then mixed  in a non-linear BBO crystal with a time-delayed reference pulse in order to generate a sum frequency signal. This last, selected with a prism and detected with a photodiode, is proportional to the temporal overlap between the signal and reference pulse. Therefore, the time profile of the transmitted signal was mapped by varying the delay of the reference (temporal resolution of 260 fs).

\begin{figure}
\includegraphics[width=\columnwidth]{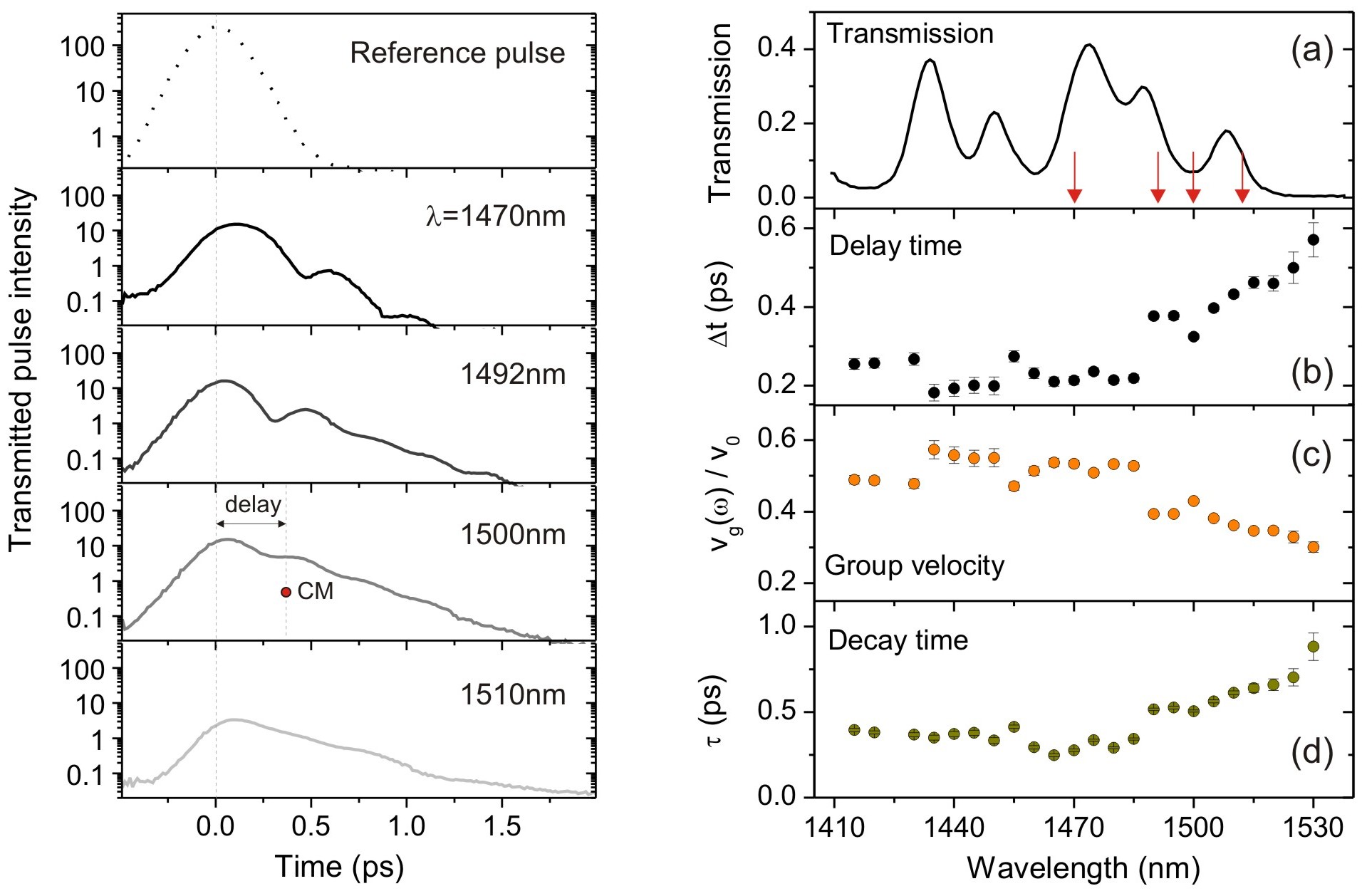}
\caption[Ultrashort pulse propagation in 1D FQs.]
{Left panels show examples of the time-resolved signals transmitted through the F12 sample. In the top panel the reference (undisturbed) pulse profile is shown, whereas the rest of the panels plot the transmitted pulses at wavelengths of 1470 nm, 1492 nm, 1500 nm, and 1510nm, respectively. The concept of the center of the mass of transmitted pulses and the corresponding delay time are explained too. Right panels report (a) the samples transmission spectrum, (b) the delay time, (c) the relative group velocity derived from the delay time and (d) the decay time of the signals as a function of the probe pulse wavelength \cite{MherFibo}.}
\label{prb2}
\end{figure}

The left panel of Fig.~\ref{prb2} reports examples of the temporal evolution of selected four transmitted pulses centered at different wavelengths of the band edge. The top panel shows the free-space pulse (with no sample) the maximum of which is taken to be the zero of the time axis. An analysis of the transmitted signals at different wavelengths provides with rich information about the nature of the band edge states;

\begin{figure}
\includegraphics[width=\columnwidth]{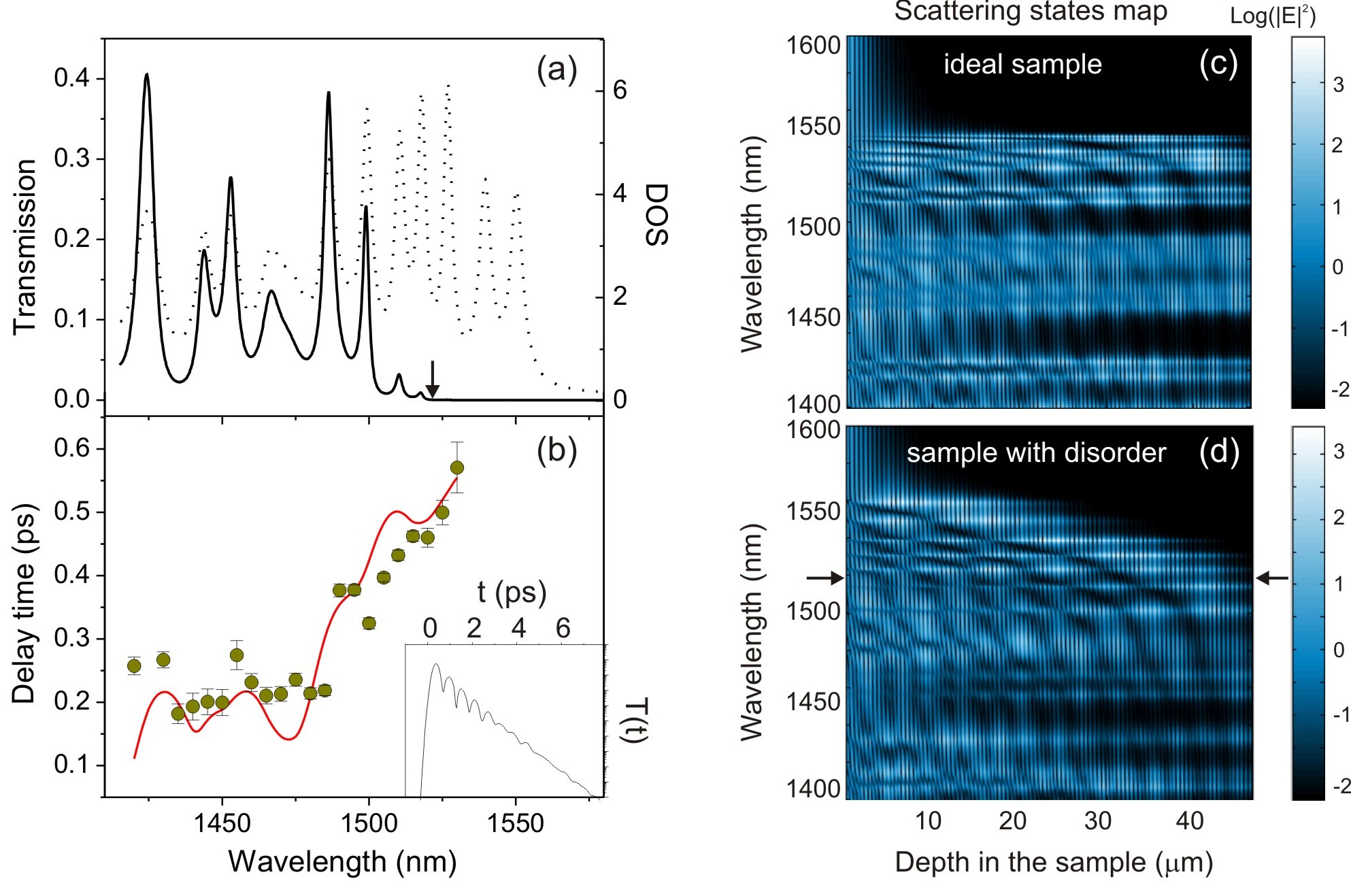}
\caption[Disorder effects on pulse propagation]
{(a) The transfer-matrix-calculated transmission spectrum (solid line)
of the F12 sample and the corresponding DOS (dotted line). (b) The experimentally measured delay times of the transmitted pulses (scatter data) are compared to the calculated ones (line). An example of the time-resolved transmitted intensity of a probe pulse centered at 1497 nm is plotted in the inset. (c) and (d) show the light intensity distribution inside the F12 sample for the ideal case (no optical path gradients) and for a 4$\%$ gradient, respectively. The arrows in the lower graph indicate the wavelength above which the sample transmission becomes almost zero, while internal modal structure can be clearly identified \cite{MherFibo}.}
\label{prb3}
\end{figure}

\emph{Disorder-related effects in 1D FQ's} -- From a quick look at the calculated DOS and the transmission shown in Fig.~\ref{prb3}(a) it is quite evident that while the first shows peaks until a wavelength of 1550 nm, the second appears to vanish already above 1520 nm. This fact indicates that there do exist states in the spectral range between 1520~nm and 1550~nm, but they barely transmit the light for some reason. The origin of this phenomenon comes from the presence of a certain degree of disorder in the quasicrystal and will be discussed in the following.

A comparison between transmission spectra, DOS, and
group velocity of a perfect DBR, an ideal Fibonacci, and a Fibonacci
structure with certain disorder has been reported and discussed in \cite{OtonFibo}. In an ideal non-absorbing structure, in which the optical path $n_i d_i$ is exactly the same for each individual layer, whenever the DOS shows a peak, it has a unity transmission. Consequently, both the DOS and $T(\omega)$ vanish in the band gap region.

The disorder inside a 1D photonic structure can be either uncorrelated to the position inside the structure or it can manifest as a continuous drift of the optical path (a linear gradient to first approximation). This last is the typical situation when during the preparation of the multilayered samples the growth rate (layer thickness) or the material composition (refractive index) is slowly varying through the final structure \cite{Gellermann,Luca,MherFibo,MherAPL}. We will therefore examine the effect of this type disorder.

When a drift is present, the low transmission regime apparently extends for a wider wavelength range than the DOS gap. The reason can be found by examining the calculated scattering states maps \index{scattering! states map} of the Fibonacci sample, reported in Fig.~\ref{prb3}(c),(d). The scattering states map for the ideal case with flat photonic band structure is shown in the top panel. The linear drift in the optical path, $d(n_i d_i)/d x$, along the depth $x$ in the multilayered structure acts as a built-in bias that tilts the band edge\index{bandgap! edge} (see Fig.~\ref{prb3}(d)). This band tilting behaves as the analogue of an external static electric field applied to an electronic \index{superlattice! electronic} superlattice\index{lattice! super-} structure. Such controlled realization of optical path gradients was exploited recently to observe time-resolved photonic Bloch oscillations \cite{Sapienza} and Zener tunneling of light in optical superlattice\index{superlattice! optical} structures with tilted photonic band structure \cite{Ghulinyan1}. Moreover, researchers have also managed to change the degree of band tilting by means of vapor flows through porous multilayers \cite{MherZenerVap,MherNatPhot}.

\begin{figure}
\includegraphics[width=\columnwidth]{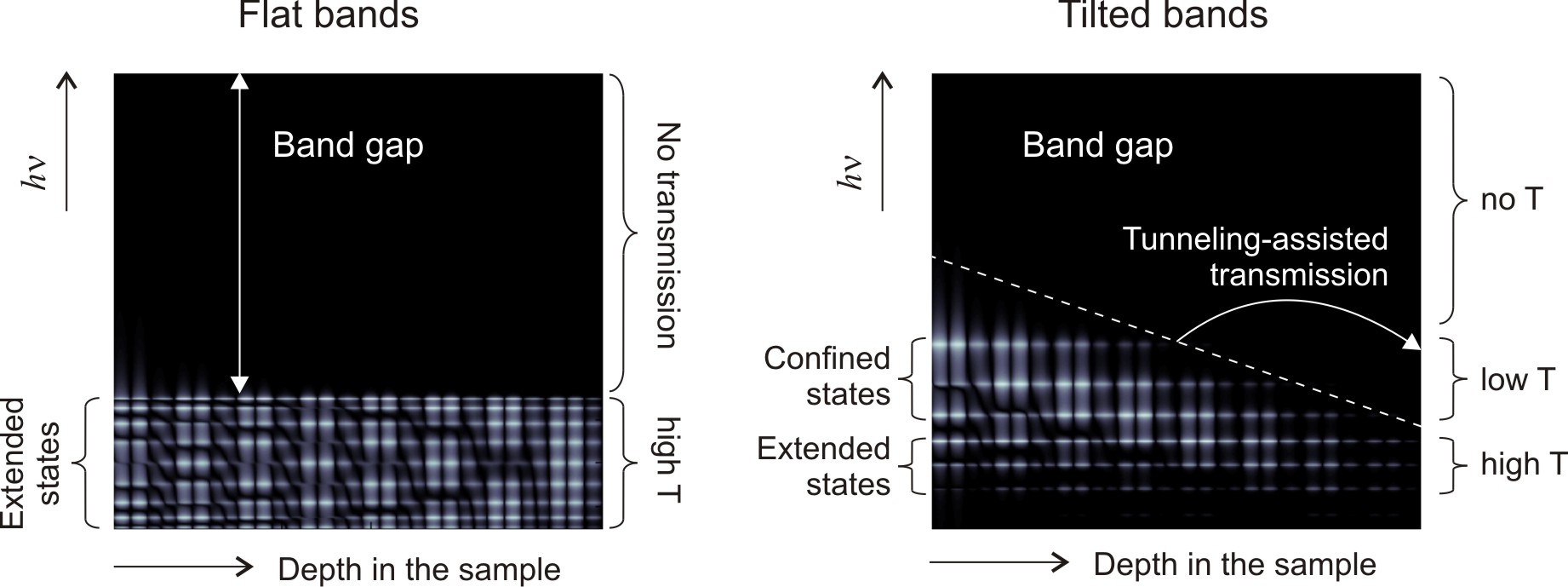}
\caption[Flat and tilted photonic bands]
{(Left graph) In the absence of optical path gradients the photonic bands are flat and the photon states are extended throughout the whole structure. (Right graph) The effect of an optical path gradient on photonic bands is in close analogy to the effect of a static electric field on energy bands in an electronic crystal. In the photonic case the band edge states are no more extended Bloch states and are confined spatially between the physical edge of the sample and the photonic band gap. Thus, photon transmission through these states is largely suppressed.}
\label{prb4}
\end{figure}

Generally, and, in particular, in the 12th order Fibonacci sample, the drift-induced band tilting reduces the spatial extension of the optical modes which are closely situated at the edge of the pseudo band gap. As a consequence, the first band edge states do not extend over the whole sample and, therefore, their contribution to high transmission channels reduces significantly. In other words, a photon may be injected easily into such a spatially squeezed state from one side of the sample, however,  in order to be transmitted it needs to tunnel through the band gap which extends for the rest of the sample (see Fig.~\ref{prb4}). This explains the fact that the transmission spectra and density of modes differ in the region of the pseudo band gap. Equivalently, the long delay times observed around the edge of the pseudo band gap can be associated with the time needed to propagate through the band edge states. The long decay times are directly related to the spatial confinement of these states.

\begin{figure}
\includegraphics[width=\columnwidth]{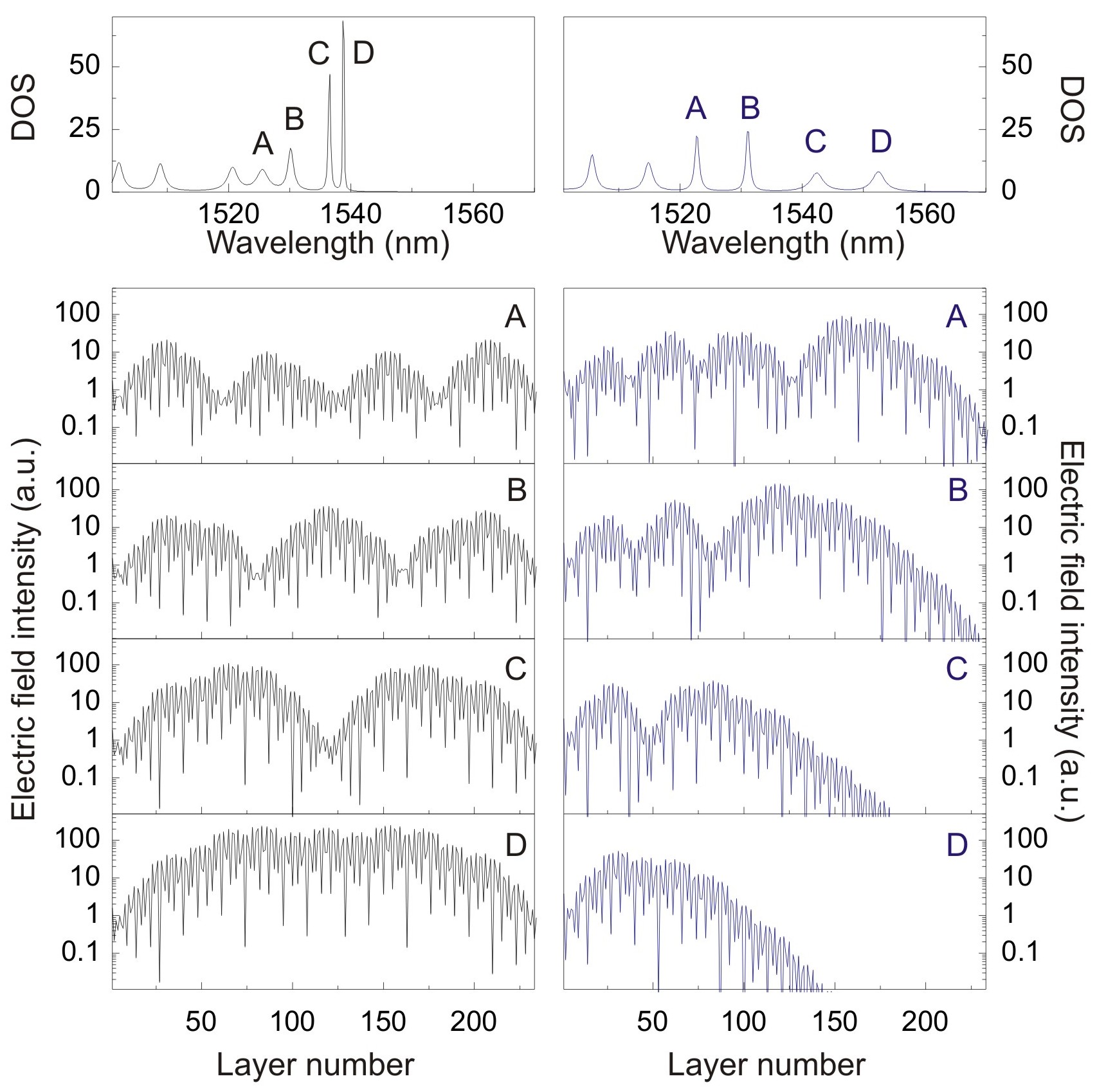}
\caption[Band edge mode profiles in an ideal FC and a sample with tilted photonic bands]
{The electric field intensity distribution inside the multilayered sample for the first four band edge states in an ideal Fibonacci structure (left panels) and a 4$\%$ drifted one (right panels). The corresponding DOS spectra around the band edge are shown in the upper panels.}
\label{efiled}
\end{figure}

Further numerical calculations in this direction have been also performed. In particular, the intensity of the electric field distribution inside the Fibonacci structure for few band edge states was investigated for both an ideal sample and one with optical path gradient. The results of these calculations are shown in Fig.~\ref{efiled} in which also the DOS at the band edge is reported in top panels. In the absence of optical path gradients (left panels), the distribution of the electric field intensity shows the characteristic self-similar structure of the band edge resonances of a Fibonacci quasicrystal.

It can be seen from the right panels of Fig.~\ref{efiled} that while the main features of the intensity distribution preserve when a drift is introduced (4\% in this case as in the experiment with F12 samples), the profiles become asymmetric with respect to the center of the sample. Quite visibly, the intensity pattern\index{pattern! intensity} squeezes more and more towards one side of the quasicrystal as the mode wavelength approaches the pseudo band gap. In accordance with the situation shown in  Fig.~\ref{prb4} for the tilted band case, the closest to the gap states are spatially more confined between the sample surface and the tilted band gap. Accordingly, when approaching the band edge (from panel A to panel B of the tilted band case), the intensity profile distorts heavily and the electric field decays efficiently through the rest of the structure. Thus, these calculations show that the natural drifts, occurring during the growth of the 1D photonic FQ, alter the spatial extension of the first few band edge modes, but do not destroy their specific structure. Higher order band edge modes are less affected by the optical path gradients.

\subsection{Origin of band edge states and pseudo band gaps}\index{localization! critical}\index{bandgap! edge} \index{bandgap! pseudo-}\index{Fibonacci! bandgap}\index{Fibonacci! quasicrystal}
In the previous subsection we reviewed the recent progress in photonic 1D Fibonacci quasicrystals in terms of their realization, spectral characterization and ultrashort pulse propagation studies. We learnt that these photonic structures are somewhat in between ideally periodic photonic crystals and disordered complex dielectric structures. When describing them, terms such as \emph{pseudo band gaps} and \emph{band edge states} with \emph{exotic (critical) localization} properties are often used. Even though, the reader might still have it unclear why ``pseudo" is used or what do we mean by saying ``critical localization" and why does it happen? The scientific literature is rather succinct in this sense and rigorous mathematical descriptions are not easy to interpret.

Through the rest of this chapter, we will offer the reader a new interpretation of these properties for the 1D photonic Fibonacci quasicrystals. We will show, in particular, how the particularity of the inflation rule of Fibonacci-type affects the properties of a 1D photonic FQ and in which way the so-called pseudo band gaps and the band edge states with non-exponential decay are formed? Perhaps, we will be rather surprised to realize that the Fibonacci quasicrystal is a particular arrangement of simple photonic building blocks -- DBR's and FP microcavities -- arranged all together to form a coupled microcavity system\index{coupled! cavities}.\\

\begin{figure}
\includegraphics[width=\columnwidth]{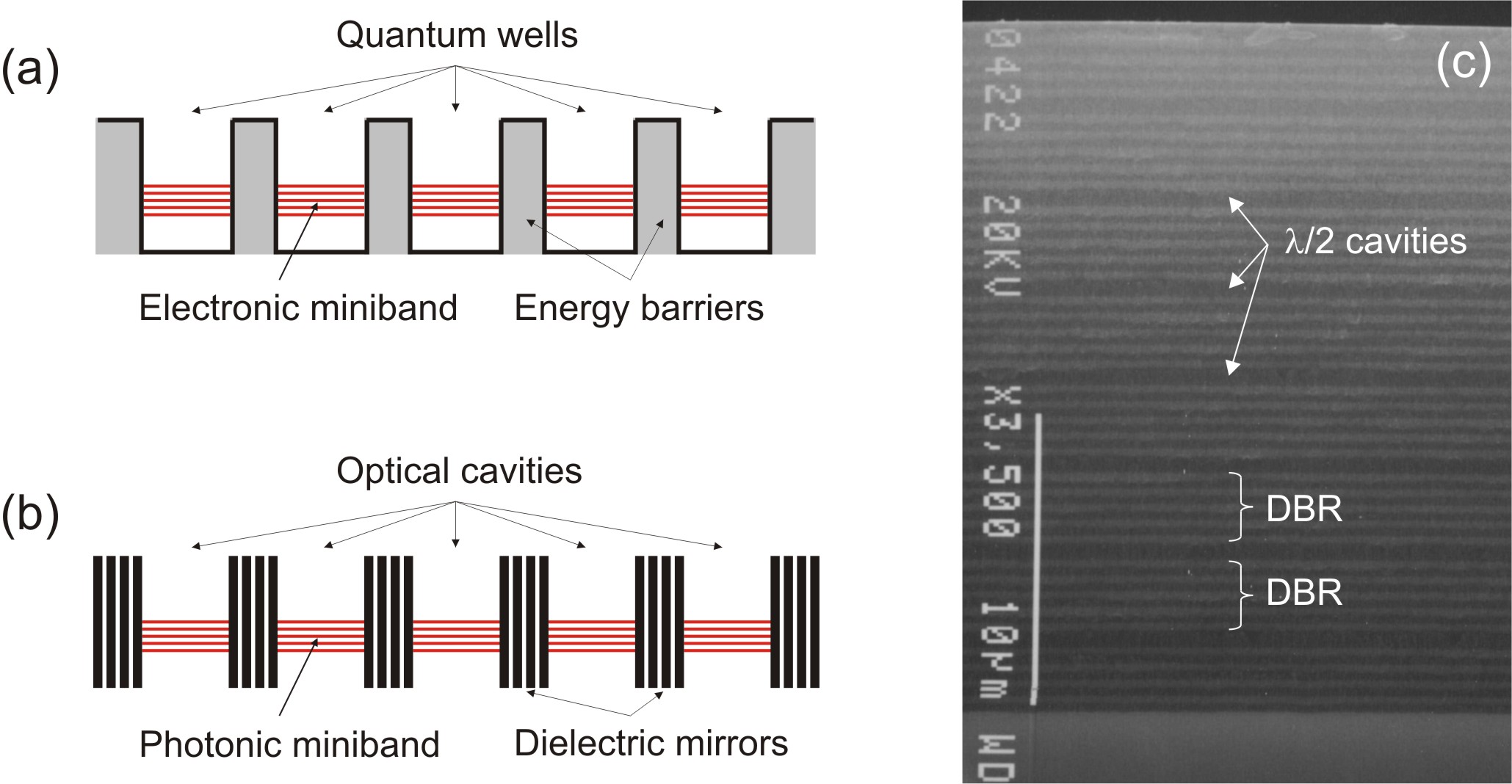}
\caption[Electronic and optical superlattices]
{In analogy to the electronic coupling of separate quantum wells in a semiconductor superlattice (a), an optical superlattice can be realized when optical cavities are brought together (b), which results in the formation of a miniband of extended photonic states. (c) A scanning electron micrograph of a one-dimensional porous silicon optical superlattice with seven coupled microcavities.}
\label{osl}
\end{figure}

\emph{Coupled microcavity optical superlattice}\index{superlattice! optical}\index{microcavity! coupled} -- In a close analogy to a semiconductor double-quantum well, the optical coupling between identical cavities lifts up the mode degeneracy inducing a repulsion, which splits the single state at a frequency $\omega_0$ into two new ones at frequencies $\omega_0\pm\Delta\omega$ \cite{pavesiPRB,CMC1,CMC2,CMC3}. Degenerate mode coupling between $N$ cavities therefore leads to the formation of a miniband\index{miniband} of $N$ new optical states, which are densely packed around the resonant frequency. This way, in a very same analogy to the electronic superlattices\index{superlattice! electronic} (Fig.~\ref{osl}(a)) \cite{esaki}, an optical superlattice can be made by coupling degenerate\index{degenerate! coupling} optical resonators\index{resonator! optical} (cavities) within the same photonic structure (Fig.~\ref{osl}(b))\cite{Shayan,Melloni,MherAPL,MherJAP,MherSlow}.

In one dimension, an optical superlattice can be realized by stacking two different quarter-wave dielectric layers $A$ and $B$ in a way to form identical cavities separated by dielectric Bragg mirrors\index{Bragg! mirror} (Fig. ~\ref{osl}(c)). In a generic form, the layer sequence of such a multilayer stack looks like [$(BA)_n B$] $(AA)_1$ [$(BA)_n B$]
$(AA)_2$ $\dots$ $(AA)_{m}$ [$(BA)_n B$], where a series of $m$ microcavities $(AA)_m$ are coupled to each other through the [$(BA)_n B$] DBRs of ($n+1/2$) periods. The amount of mode splitting of cavity\index{cavity! resonance splitting} resonances is given by the strength of the coupling mirrors. As described in the Section~\ref{1dphot}, for a given refractive index contrast, the reflectance of the DBR improves with increasing the number of periods. Similarly, for a fixed number of periods, the mirror reflectance grows with an increase in the refractive index contrast, i.e. the stronger the mirrors reflect, the weaker is the mode repulsion.

\begin{figure}
\includegraphics[width=\columnwidth]{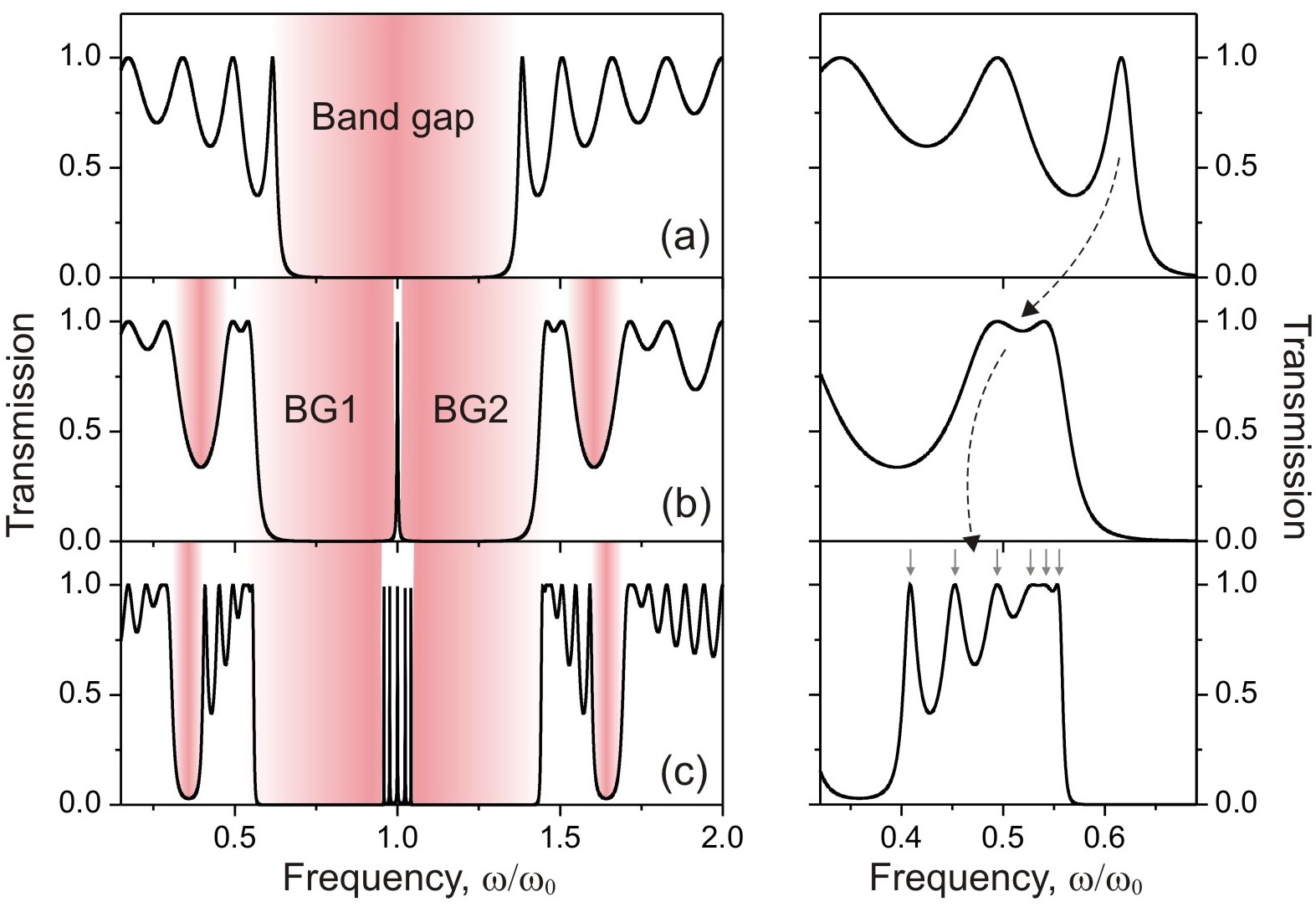}
\caption[From single to multiple cavity photonic crystals]
{Broad range transmission spectra of (a) a dielectric mirror, (b) a Fabry-P$\acute{e}$rot microcavity and (c) five coupled microcavities. The zooms of spectra around the band edge states are shown in relative panels in the right.}
\label{cmc}
\end{figure}

Figure~\ref{cmc} shows the formation of a characteristic transmission spectrum of an optical superlattice of five coupled microcavities. First, let us look at the spectrum of a single DBR which has a layer sequence [$(BA)_n B$] with $n=5$ (top panel)\footnote{The exact number of DBR periods $n$ is not essential throughout this analysis.}. As discussed previously it is characterized by a photonic band gap centered at a relative frequency $\omega_0/\omega=1$ and extends to a certain frequency range around it (indicated by the shaded area).

Next, we introduce a defect in the DBR layer sequence creating thus a $\lambda/2$ cavity sandwiched between two DBR's: this creates a sharp transmission peak -- the Fabry-P$\acute{e}$rot resonance -- around the central frequency and splits the DBR band gap in two forbidden spectral regions (Fig.~\ref{cmc}(b)). It is important to underline that the new band gaps on the left and right sides of the FP peak are part of the \emph{same and unique} band gap of the DBR sequence [$BABA\ldots B$] and therefore have the center frequency $\omega_0/\omega$.

Moreover, the FP-resonators's\index{resonator! Fabry P$\acute{e}$rot} spectrum contains other details which are absent in the DBR's spectrum. While these details are not evident at a first glance, they can be revealed by looking at the blow up of the corresponding $T(\omega)$ curves in the vicinity of the band edge. From the right panels of Fig.~\ref{cmc} we notice, in particular, that with respect to the DBR's spectrum the band edge states of the FP structure have undergone spectral reshaping. Namely, a paired (double-peaked) band edge state is appearing now. Understanding the origin of these doublet is important as it will be a crucial point when analyzing the origin of Fibonacci pseudo band gaps.

Recalling the layer sequence of the 1D FP microcavity:
\begin{center}
 [$BABA\ldots B$] $AA$ [$BABA\ldots B$]
\end{center}
and remembering that by definition they have the same central frequency, we notice that two identical DBR's are separated through a layer $AA$, which can be interpreted as a system of coupled DBR's\index{coupled! dielectric mirrors}! As already discussed above, such a coupling should induce degenerate mode splitting\index{degenerate! mode splitting}. Thus the origin of the band edge doublet comes from a coupling of the sandwiching DBR's in the FP structure sequence. This coupling is valid for all DBR states. In fact, by looking back at broad range transmission spectrum of the FP microcavity (Fig.~\ref{cmc}(b)), we notice that the doublets are present for all other modes far from from the band edge. Moreover, the formation of these doublets induces relatively deep low-transmission (gap-like) regions between neighboring duplets. The successive gap-like regions become shallower when moving further from the band edge (shallower shaded areas in Fig.~\ref{cmc}(b)).

These observations become more evident when looking at the spectrum of a five coupled microcavity structure (Fig.~\ref{cmc}(c)). First, we notice the formation of photonic miniband states, five in total, which are densely packed around $\omega_0$. As in the case of the single FP-structure, this miniband is separating the fundamental DBR band gap\index{bandgap! fundamental} in two regions at $\omega<\omega_0$ and $\omega>\omega_0$. Secondly, from corresponding zoom of the band gap edge we notice that now we have a multi-peaked modal structure (right bottom panel). Importantly, one counts six peaks which correspond exactly to the total number of DBRs in a five coupled microcavity structure. Moreover, the secondary gap-like features become much deeper and well defined. Finally, these observation confirm the fact that in analogy to the coupling between identical $\lambda/2$ $AA$ layers also the dielectric mirrors couple within the same photonic crystal.

It is now the right moment to look at the formation of the band structure of a 1D photonic Fibonacci quasicrystal.\\

\emph{1D photonic Fibonacci quasicrystal} -- Let us first examine the layer sequence of the FQ. As an example we consider the 7th order Fibonacci sequence, which reads
\begin{center}
$ABAABABAABAABABAABABA$.
\end{center}
Now, remembering that in the photonic case (i) $A$ and $B$ layers have a  quarter-wave thickness and, hence, (ii) an $AA$ layer forms a $\lambda/2$ cavity, we can write this Fibonacci sequence in the form
\begin{center}
[$AB$] $AA$ [$BAB$] $AA$ [$B$] $AA$ [$BAB$] $AA$ [$BABA$].
\end{center}

\begin{figure}
\includegraphics[width=\columnwidth]{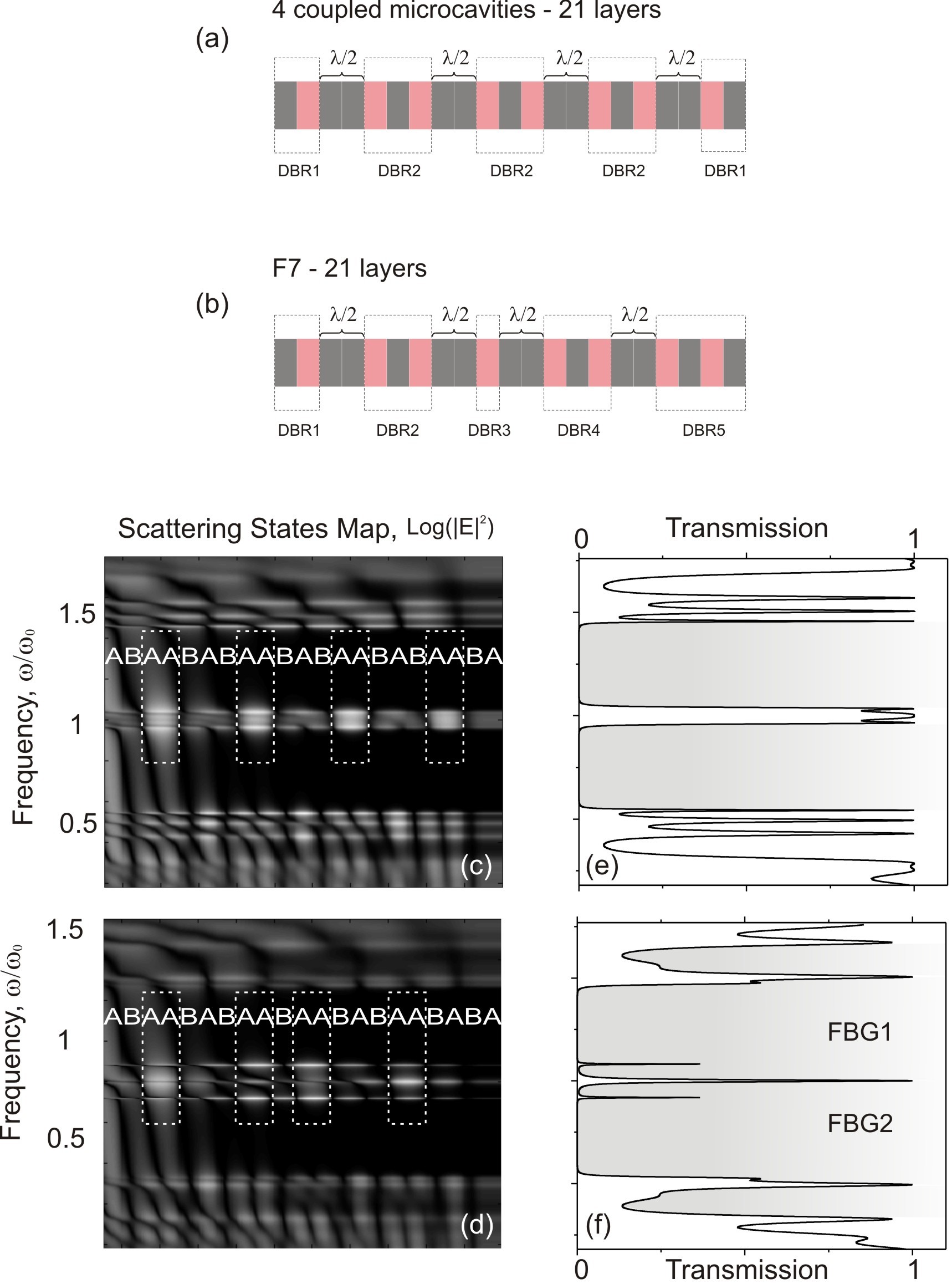}
\caption[A Fibonacci quasicrystal vs a coupled microcavity photonic crystal]
{(a), (b) The schematics of layer alternation in a four coupled microcavity and in a 7th order FQ structure. (c) The scattering states map and (e) the corresponding transmission spectrum of the coupled microcavity structure. (d) and (f) show the same quantities for the 1D FQ. }
\label{ssmFiboCMC}
\end{figure}

We notice immediately, that this Fibonacci multilayer\index{Fibonacci! multilayer} is nothing else than \textbf{a system of four coupled microcavities with unequal dielectric mirrors} (see Fig.~\ref{ssmFiboCMC}(a) and (b)). It is important to note here again the fact that due to the recurrence rule of the Fibonacci type, the $AA$ is a frequent instance along an $F_n$ for large $n$'s, while other instances such as $BB$ or $AAA$ may never appear. In view of this, we can say now that a 1D photonic Fibonacci quasicrystal, independently on its recurrence order $n$, will be a coupled $\lambda/2$-microcavity system with unbalanced DBR's.\footnote{In order to form coupled cavities (three at least, since two cannot be formed), $n$ should be larger than 6. F4 and F5 have only one cavity, while no cavity is formed in lower orders F1, F2, and F3.}

\begin{table*}
\caption[Coupled cavities in Fibonacci sequences]
{\small{The occurrence of $AA$-cavities in layer sequences of 1D FQs (from 3rd to 7th Fibonacci orders).}}
\label{sample-table}
\addtolength\tabcolsep{2pt}
\begin{tabular}{@{}c@{\hspace{25pt}}ccc@{}}
\hline \hline
\raggedleft
1D FQ & counts($A$) & counts($B$) & counts($AA$)\\
\hline
\multicolumn{1}{|l|} {$ABA$} & 2 & 1 & 0\\[3pt]
\multicolumn{1}{|l|} {$AB|AA|B$} & 3 & 2 & 1\\[3pt]
\multicolumn{1}{|l|} {$AB|AA|BABA$} & 5 & 3 & 1\\[3pt]
\multicolumn{1}{|l|} {$AB|AA|BAB|AA|B|AA|B$} & 8 & 5 & 3\\[3pt]
\multicolumn{1}{|l|} {$AB|AA|BAB|AA|B|AA|BAB|AA|BABA$} & 13 & 8 & 4 \\[3pt]
\hline \hline
\end{tabular}
\end{table*}

For comparison, in Fig.~\ref{ssmFiboCMC}(c),(d) the light intensity distributions inside a four coupled microcavity (4CMC) structure with balanced DBR's \cite{MherSlow} and an F7 1D photonic quasicrystal are shown. Their respective transmission spectra are plotted in Fig.~\ref{ssmFiboCMC}(e),(f). We note that the weaker external mirrors in the case of the 4CMC structure are chosen intentionally such to count a total of 21 layers as in a F7 structure. It is however important that the intra-cavity (coupling) mirrors are all identical. These intensity maps visualize nicely the layer positions of the $\lambda/2$-cavities. In particular, one can observe that in the case of 4CMC structure they are evenly distributed through the multilayer (see the bright intensity spots around $\omega/\omega_0=1$) in contrast to their inhomogeneous distribution in an F7 quasicrystal.

It follows thus that the optical coupling of cavities inside a Fibonacci multilayer is depth-dependent. Also, it is clear that this coupling is not changing in a random manner with the position inside the multilayer, but is related to the exact DBR sequences which strictly reflect the Fibonacci string generation rule. In other words, in a 1D FQ the sequences of the DBRs are fixed by the recurrence rule. For example, the first five DBR's are $AB$, $BAB$, $B$, $BAB$ and $BABA$.

In view of the above discussion certain aspects of FQ \emph{pseudo band gap} formation and the nature of the \emph{band edge states} are seen under a new light. In particular,
\begin{itemize}
  \item The so-called fundamental pseudo band gaps (FBG1 and FBG2 in Fig.~\ref{ssmFiboCMC}(f)) are essentially parts of a unique band gap separated by the photonic miniband\index{miniband} of coupled cavities,
  \item Higher order pseudo band gaps are low-transmission regions formed by neighboring multi-peaked states of the coupled DBRs (more external shaded regions in Fig.~\ref{ssmFiboCMC}(f)). The origin of their formation is exactly the same as in the case of coupled microcavity structures with balanced mirrors (see Fig.~\ref{cmc}(c)).
  \item Fundamental band edge states are of two different types. The first are coupled cavity miniband states and originate from $AA$ layers. The band edge states of second type are essentially coupled DBR states and with respect to the coupled cavity states are situated on the opposite side of the gap.
\end{itemize}

We now examine the properties of the band edge states of the first type, i.e. the $AA$-layers-related coupled cavity\index{cavity! coupled} miniband of the Fibonacci quasicrystal. As already mentioned above, the coupling of cavities inside the 1D FQ takes place via unbalanced mirrors. This results in the formation of photonic states which are not homogeneously distributed within the spectral width of the miniband, as opposed to the case of a coupled microcavity system with uniform DBRs \cite{MherAPL,MherJAP}. For a more detailed analysis we consider an 8th order FQ which counts eight occurrences of $AA$-cavities. For simplicity, we rename the $AA$-cavities to $C_m$ ($m$=1,2,...,8) and rewrite the Fibonacci sequence as
\begin{center}
$AB$ $C_1$ $BAB$ $C_2$ $B$ $C_3$ $BAB$ $C_4$ $BAB$ $C_5$ $B$ $C_6$ $BAB$ $C_7$ $B$ $C_8$ $B$.
\end{center}

\begin{figure}
\includegraphics[width=\columnwidth]{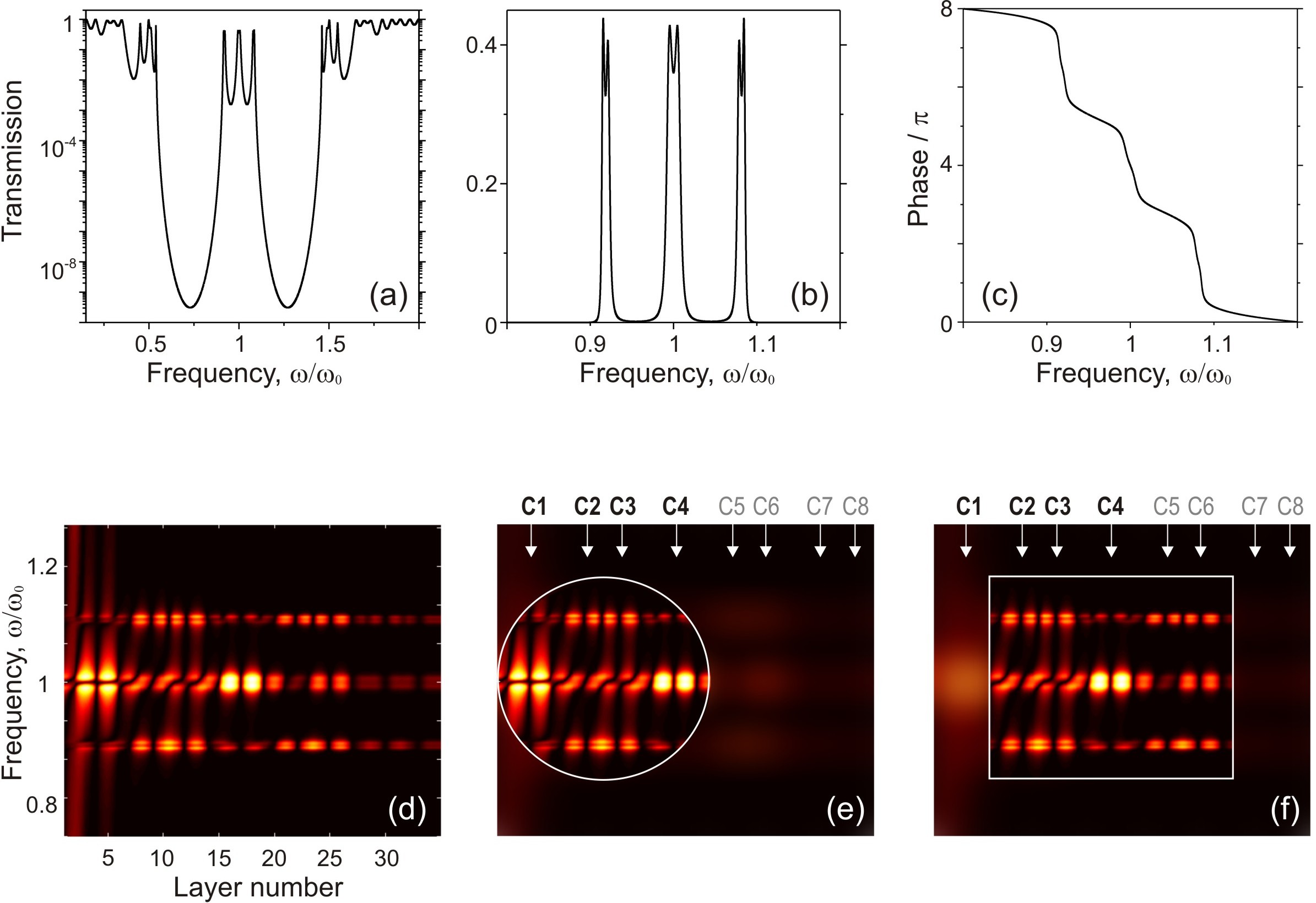}
\caption[Nonuniform mode splitting in a photonic Fibonacci quasicrystal]
{(a) The broad range transmission spectrum of an 8th order FQ. (b) The zoom of the spectrum and (c) the accumulated phase around the miniband region. (d) The scattering states map of the miniband states. Panels (e) and (f) explain the degenerate mode coupling between various cavities distributed through the Fibonacci sequence.}
\label{MB}
\end{figure}

The typical broad range transmission spectrum as well as the zoom of the miniband region are plotted respectively in panels (a) and (b) of Fig.~\ref{MB}. As we can see, the miniband is composed of three regions of relatively high transmission states. While a total of six marked peaks can be counted, the number of states involved in the miniband is eight whereby two states are less pronounced and are not visible in the $T(\omega)$ spectrum. The accumulated phase, which is a very useful tool when analyzing unknown resonant structures  \cite{JacPRE,MherNeck1,MherNeck2,MherSlow}, is plotted in Fig.~\ref{MB}(c) and counts a total of exactly $8\pi$ when sweeping the frequency through the miniband region. The three high-$T$ channels are separated by two regions of very low transmission. These last are often attributed to the ``category" of higher order pseudo band gaps, while their origin is clearly related to the cavity miniband structure.

In order to understand better the formation of the inhomogeneously split miniband we refer to the electric field intensity distribution inside the quasicrystal around the miniband central frequency $\omega_0$ (Fig.~\ref{MB}(d)). The positions of the bright spots, which indicate to the large filed intensities accumulated in cavities, show certain symmetric patterns both through the depth inside the structure ($x$-axis) as well as over the frequency range ($y$-axis) around the $\omega/\omega_0=1$ point.

As an example, we analyze two of the symmetric patterns, which are shown in panels (e) and (f) of Fig.~\ref{MB}. First, we notice that cavities $C_1$ and $C_4$ are weakly coupled through the large mirror $BABC_2BC_3BAB$, therefore the resulting mode splitting is small (on $y$-axis). More precisely, this double-peaked state is formed by a symmetric coupling of three cavities $C_1$, $C_4$ and $C_7$ since an identical mirror $BABC_5BC_6BAB$ is present between $C_4$ and $C_7$:
\begin{center}
$AB$ $C_1$ [$BAB\,C_2\,B\,C_3\,BAB$] $C_4$ [$BAB\,C_5\,B\,C_6\,BAB$] $C_7$ $B$ $C_8$ $B$.
\end{center}
We notice also that the cavities $C_2$ and $C_3$, which are strongly coupled through the very weak mirror $B$, therefore, the resulting mode splitting pushes the new states far from $\omega/\omega_0=1$ to the extremes of the miniband \cite{MherNeck1,MherNeck2}. Moreover, each of these new states itself is a doublet state, which can be explained by further analyzing the FQ structure. Namely, as in the case of $C_2$-$C_3$ coupling, we notice that the same scenario of coupling takes place also for $C_5$ and $C_6$ cavities, which pushes the new split states exactly to the same frequencies as for the $C_2$-$C_3$ case forming thus a new couple of degenerate states at the extremes of the miniband. In this case, however, the coupling between cavity pairs $C_2$-$C_3$ and $C_5$-$C_6$ is much weaker;
\begin{center}
$AB$ $C_1$ $BAB$ $C_2\,B\,C_3$ [$BAB\,C_4\,BAB$] $C_5\,B\,C_6$ $BAB$ $C_7$ $B$ $C_8$ $B$,
\end{center}
therefore, only a small amount of mode splitting is observed.\\

Thus, while this kind of analysis can be done further considering other cavity pairings (we leave the reader to exercise with), the formation of inhomogeneously distributed miniband states in a 1D photonic quasicrystal of Fibonacci type is comprehensible now.\\

\subsection{Localisation properties of band edge states}
\begin{figure}[h!]
\includegraphics[width=\columnwidth]{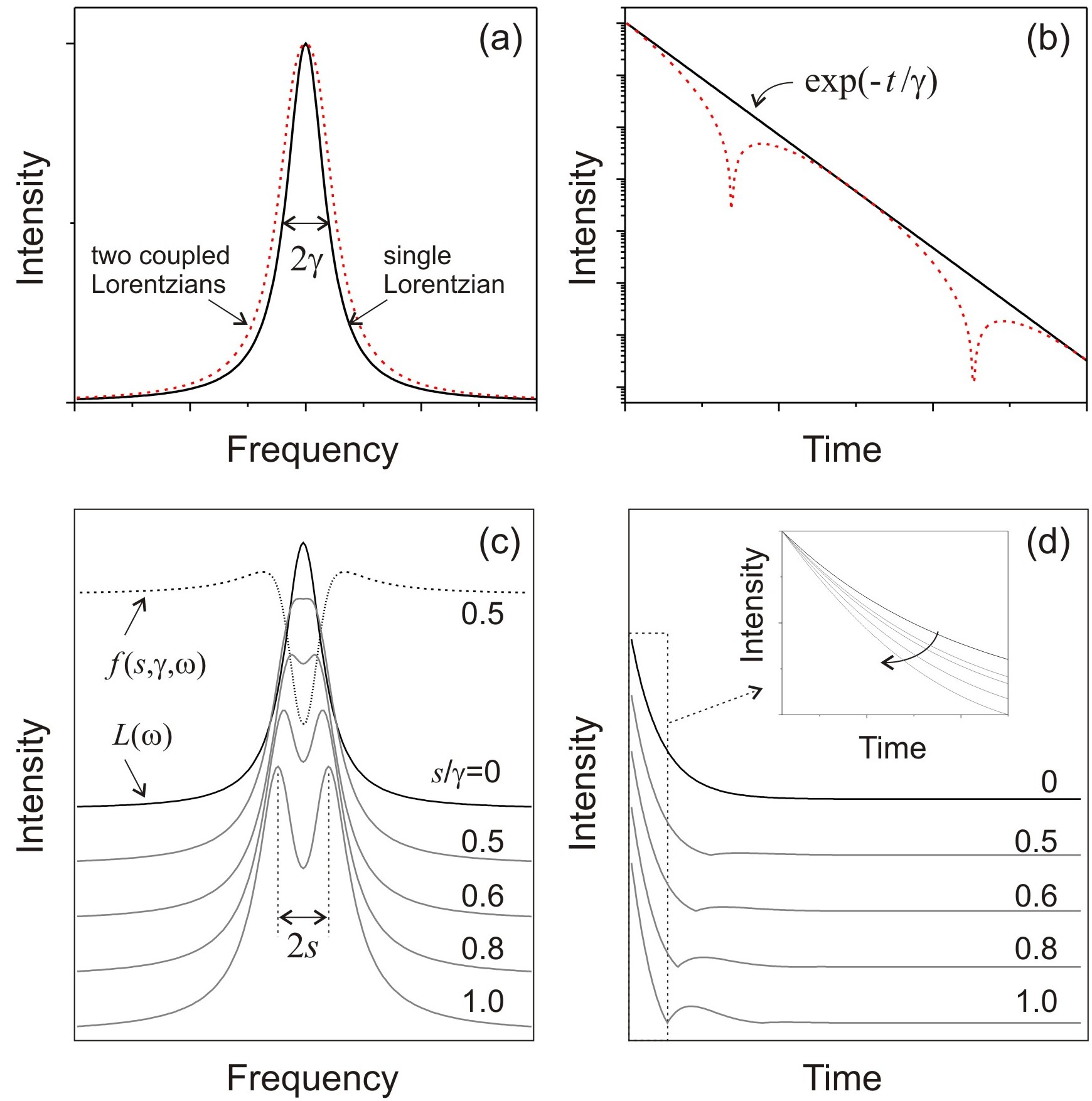}
\caption[The comparison of the spectral lineshape and the decay of a single resonance and two coupled resonances.]
{(a) Spectral lineshape of a single (solid line) and two weakly coupled Lorentzian (dashed line). (b) The corresponding inverse FFT transformations of the single and coupled states. (c) The evolution of the doublet splitting with increasing coupling, $s/\gamma$. (d). The corresponding time decays of doublet-states showing the oscillatory-time behavior. The insets is the zoom of the time-spectra at the initial part of the decays.}
\label{Lor_decay}
\end{figure}

It is well-known that the energy spectrum of a localized at a frequency $\omega_0$ state is described by a complex function
\begin{equation}
t(\omega)\sim\frac{\gamma}{\omega-\omega_0 + i \gamma},
\label{single_resonance}
\end{equation}
and its transmission spectrum has a Lorentzian lineshape (Fig.~\ref{Lor_decay}(a))
\begin{equation}
L(\omega)=|t(\omega)|^2\sim\frac{\gamma^2}{(\omega-\omega_0)^2 +\gamma^2}.
\label{single_Lor}
\end{equation}
Here, $2\gamma$ is the full width at the half-maximum (FWHM) of the resonance. In other words, $\gamma$ describes how good the cavity confines the electromagnetic energy or, equivalently, how lossy it is. Thus, the FWHM in the frequency-domain represents the decay rate, proportional to $\gamma^{-1}$, of the resonance in the time-domain. In fact, the inverse Fourier transformation of a Lorentzian lineshape is an exponentially decaying in time function (solid line in Fig.~\ref{Lor_decay}(b)).

In case when two identical resonances (same $\gamma$) are degenerately coupled, the resulting lineshape is described through a coherent sum of two Lorentzians as
\begin{equation}
D(\omega)\sim\left|\frac{\gamma}{\omega-\omega_0+s +i \gamma}+\frac{\gamma}{\omega-\omega_0-s +i \gamma}\right|^2,
\label{double_Lor}
\end{equation}
where 2$s$ is the coupling induced splitting of the doublet resonance. This lineshape can be also presented as
\begin{equation}
\begin{aligned}
D(\omega)\sim {} & \left|\frac{\gamma}{\omega-\omega_0-i \gamma}\right|^2\times \\
& \left|\frac{2\left[(\omega-\omega_0)^2+\gamma^2\right]}{(\omega-\omega_0)^2-\gamma^2-s^2 +2 i \gamma(\omega-\omega_0)}\right|^2,
\label{double_Lor2}
\end{aligned}
\end{equation}
This result has the form $L(\omega)\times f(s,\gamma,\omega)$, in which we recognize the single-resonance Lorentzian $L(\omega)$ convoluted with the function $f(s,\gamma,\omega)$. It is easy to see that $\lim \limits_{s \to 0} D(\omega)=L(\omega)$, i.e. when the coupling is infinitely weak, one ends up with a single Lorentzian lineshape.

In a weak coupling situation (e.g., a strong intracavity mirror) the splitting is small and a clear doublet is not formed (dotted line in Fig.~\ref{Lor_decay}(a)). While this lineshape appears as an effectively broader ``single-peaked" resonance without any evident fine spectral features, its Fourier spectrum reveals a clearly faster decay (dotted line in Fig.~\ref{Lor_decay}(b)) than the pure exponential one of the Lorentzian resonance. Moreover, the composite nature of this double-state resonance is brought to light nicely in the time-domain; it can be observed at a relatively longer time scale that the state decay is not monotonic but it performs regular oscillations. The time period of these oscillations is defined by the inverse of the doublet splitting, $s^{-1}$. This is what typically is observed in time-resolved pulse propagation experiments \cite{Luca,MherFibo,Sapienza,Ghulinyan1}, which essentially reflects the convolution of the ultrashort pulse lineshape with that of the composite resonance; regular beatings occur between two coupled states which manifest as oscillations in the transmitted intensity (see Fig.~\ref{prb2}). In addition, the transmitted pulse is also characterized by a larger
delay which is caused by the transient time necessary to build up
the double resonance.

Figure~\ref{Lor_decay}(c) shows the evolution of the spectral lineshape of two coupled Lorentzian states at increasing coupling strength $s/\gamma$. The corresponding iFFT spectra, plotted in Fig.~\ref{Lor_decay}(c), show that the oscillations become faster for larger doublet splittings and the intensity decays faster at short times (see the inset of Fig.~\ref{Lor_decay}(c)).

When more than two resonances are degenerately coupled within a photonic structure, the decay of such a state may show an even more complicated time-behavior. In particular, more than one oscillation period can appear in the intensity decay profile, but as well the amplitude of beatings can vary because of unevenly distributed degenerate resonances within the structure. A photonic structure which fits well this situation is the 1D Fibonacci quasicrystal. As we discussed earlier in this subsection, the band edge states of a 1D FQ are miniband states originating from $AA$ half-wavelength layers unevenly distributed throughout the photonic quasicrystal. While for their localization properties terms such as ``exotic" or quasi-localization are typically used to describe a weaker-than-exponential decay, we see that this property is exactly what is expected for photonic resonances originating from degenerately coupled states.

\section{Acknowledgments}
We wish to thank our co-workers Claudio J. Oton, Luca Dal Negro, Lorenzo Pavesi, Riccardo Sapienza, Marcello Colocci, Diederik S. Wiersma for their important contributions to the study of 1D Fibonacci quasicrystals presented in this chapter. We are also thankful to Daniel Navarro-Urrios, Paolo Bettotti, Zeno Gaburro, Georg Pucker, Matteo Galli, Lucio C. Andreani and Andrea Melloni for many fruitful discussions and our common studies on 1D coupled photonic microresonator systems. This work was financially supported by the INFM projects RANDS and Photonic and by MIUR through the Cofin 2002 ``Silicon based photonic crystals" and FIRB ``Sistemi Miniaturizzati per Elettronica e Fotonica" and ``Nanostrutture
molecolari ibride organiche-inorganiche per fotonica" projects.

%

\end{document}